\documentclass[structabstract]{aa}

\usepackage{graphicx}
\usepackage[intlimits]{amsmath}
\usepackage{txfonts}
\usepackage{natbib}
\usepackage{color}
\bibpunct{(}{)}{;}{a}{}{,}

\begin{document}
\title{Treating gravity in thin-disk simulations}

\author{
	Tobias~W.~A.~M\"uller\inst{\ref{inst1}}
	\and Wilhelm~Kley\inst{\ref{inst1}}
	\and Farzana~Meru\inst{\ref{inst1}, \ref{inst2}}
}

\institute{
	Institut f\"ur Astronomie \& Astrophysik, 
	Universit\"at T\"ubingen,
	Auf der Morgenstelle 10,
	72076 T\"ubingen,
	Germany
	\label{inst1}
\and
	Institute for Astronomy,
	ETH Z\"urich, 
	Wolfgang-Pauli-Strasse 27,
	8093 Z\"urich,
	Switzerland
	\label{inst2}
}

\date{Received 23 December 2011 / Accepted 6 March 2012}

\abstract
{In 2D-simulations of thin gaseous disks with embedded planets or self-gravity the gravitational potential needs to be smoothed
to avoid singularities in the numerical evaluation of the gravitational potential or force.
The softening prescription used in 2D needs to be adjusted properly to correctly resemble the realistic case of vertically extended 3D disks.}
{We analyze the embedded planet and the self-gravity case and provide a method to evaluate the
required smoothing in 2D simulations of thin disks.}
{Starting from the averaged hydrodynamic equations and using a vertically isothermal disk model,
we calculate the force to be used in 2D simulations. We compared our results
to the often used Plummer form of the potential, which runs as $\propto 1/(r^2 + \epsilon^2)^{1/2}$.
For that purpose we computed the required smoothing length $\epsilon$ as a function of distance $r$ to the planet
or to a disk element within a self-gravitating disk.}
{We find that for longer distances $\epsilon$ is determined solely by the vertical disk thickness $H$.
For the planet case we find that outside $r \approx H$ a value of $\epsilon = 0.7 H$ describes the averaged force very well,
while in the self-gravitating disk the value needs to be higher, $\epsilon = 1.2 H$. 
For shorter distances the smoothing needs to be reduced significantly. 
Comparing torque densities of 3D and 2D simulations we show that the modification to the vertical density stratification
as induced by an embedded planet needs to be taken into account to obtain agreeing results.}
{It is very important to use the correct value of $ \epsilon $ in 2D simulations to obtain a realistic outcome.
In disk fragmentation simulations the choice of $ \epsilon $ can determine whether a disk will fragment or not.
Because a wrong smoothing length can change even the direction of migration, it is very important to
include the effect of the planet on the local scale height in 2D planet-disk simulations.
We provide an approximate and fast method for this purpose that agrees very well with full 3D simulations.}

\maketitle

\section{Introduction}
\label{sec:introduction}

Numerical simulations of accretion disks are often performed in the two-dimensional (2D) thin-disk
approximation, because a full 3D treatment with high resolution is still a computationally demanding endeavor
requiring a lot of patience.
In numerical disk models with embedded planets and/or self-gravity the gravitational potential (force) needs to be smoothed
because it diverges for very short mutual distances. For that purpose the potential is softened 
to avoid the singularities that arise, for example, from pointlike objects, such as planets embedded in disks.
In full 3D simulations this smoothing may be required solely for stability purposes and can be chosen to be as small as the
given numerical resolution allows. In contrast, 2D disk simulations are typically based on a vertical averaging procedure
that leads to a {\it physically required} smoothing. Ideally, this smoothing should be calculated in a way that the
2D simulations mimic the 3D case as closely as possible.
The most often used potential smoothing has a Plummer form with $\Psi \propto - 1/(r^2 + \epsilon^2)^{1/2}$,
where $r$ is the distance to the gravitating
object and $\epsilon$ is a suitably chosen smoothing or softening length.
If a vertically stratified disk has a thickness $H$, we would expect
that somehow $\epsilon$ should depend suitably on $H$. In a sense, the potential is `diluted' in this case due
to the disk's finite thickness.

For the planet-disk problem, \citet{1999ApJ...516..451M} have shown with local shearing sheet simulations owing
due to this dilution effect, the total torque exerted on a planet in a three dimensional disk is only about 43\% of the
torque obtained in the thin, unsmoothed 2D case. These authors also showed that the strength of the one-sided torque depends on the value of
the smoothing, where larger $\epsilon$ lead to smaller torques.
Later, \citet{2002A&A...387..605M} has studied the smoothing problem in greater detail. He has shown that good agreement of
the total 2D and 3D Lindblad torques can be obtained for smoothing lengths of $\epsilon = 0.75$H, where $H$ is the vertical
scale height of the disk, see Eq.~(\ref{eq:height}) below. \citeauthor{2002A&A...387..605M} found that for the Lindblad torques this optimum
$\epsilon/H$ value is independent of the planet mass and the thickness of the disk. 
On the contrary, for the corotation torques that are generated by material moving on horseshoe orbits, he found that
the required smoothing depends on the ratio $R_{\rm H}/H$, where $R_{\rm H}$ is the Hill radius of the planet.
\citeauthor{2002A&A...387..605M} concluded that there is no `magic' value of $\epsilon$ that generates overall agreement of 2D with 3D results.
  
Based on these studies, in 2D planet-disk simulations the parameter $\epsilon$ is typically chosen such that the {\it total} 
torque acting on the planet, which determines the important migration speed,
is approximately equal to that obtained through 3D (linear) analysis, for example by \citet{2002ApJ...565.1257T}.
This argument has led to the choice of $\epsilon \approx 0.3-0.6 H$, a value very often used in these simulations
\citep{2002A&A...387..605M,2006MNRAS.370..529D,2009MNRAS.394.2283P}.
However, recently a very small smoothing has been advocated for 2D planet-disk simulation \citep{2011ApJ...741...56D}.
A smoothing based on a vertical integration using Gaussian density profiles has been used by 
\citet{2005ApJ...624.1003L,2009ApJ...690L..52L}, but they provided no details on the methodology and accuracy.

For self-gravitating disks, the conditions for fragmentation have recently attracted much attention
in the context of planet formation via gravitational instability.
Typically, studies are performed in full 3D (see e.g. \citet{2011MNRAS.410..559M} and references therein). 
However, to save computational effort, 2D simulations present an interesting alternative in
this context \citep{2011MNRAS.416L..65P}.
Here, the accurate treatment of self-gravity is very important.

The incorporation of self-gravity in 2D thin-disk calculations can be achieved by fast Fourier transforms.
\citet{2008ApJ...678..483B} presented a method where the force is calculated directly using a smoothing length
that scales linearly with radius, and they used the same smoothing, $\epsilon = 0.3H$, for the planet and self-gravity.
\citet{2005ApJ...624.1003L} used this method to calculate the potential in the disk's midplane.
The simultaneous treatment of an embedded planet and disk-self-gravity can be important because
the latter may influence the migration properties of the planet \citep{2005A&A...433L..37P,2008ApJ...678..483B}. 
For global self-gravitating disks the treatment of the potential has been analyzed in more detail by \citet{2009A&A...507..573H},
who calculated the required smoothing by comparing 3D and 2D disk models with specified density stratifications.
For that purpose they compared the midplane value of the 3D potential to the 2D case and estimated from this
the required smoothing length. Additionally, they considered the whole extent of the disk for their integration.
They found that $\epsilon$ needs to be reduced for close separations $\approx H$, while for long distances it
approaches a finite value. \citet{2009A&A...507..573H} give an extended list of smoothing prescriptions used in the
literature and we refer the reader to their paper.

Here, we will reanalyze the required smoothing in 2D disk simulations. We show that the force to be used in 2D 
has to be obtained by performing suitable vertical averages of the force. 
For the planet-disk case we extend the work by \citet{2002A&A...387..605M} and
compare in detail the torque density to full 3D simulations. We present a method to approximately
include  the change in vertical stratification induced by the presence of the planet. 
With respect to self-gravitating disks we follow the work by \citet{2009A&A...507..573H} and calculate the optimum smoothing length
$\epsilon$ by performing a vertical averaging procedure for the force between two disk elements.
We show that this will be important for fragmentation of gravitational unstable disks.

In the next two sections we present the vertical averaging procedure and our unperturbed isothermal disk model.
In Sect. \ref{sect:planet} we analyze the potential of an embedded planet followed by the self-gravitating case.
In Sect. \ref{sect:summary} we summarize and conclude.

\section{The vertical averaging procedure}
\label{sec:average}

Throughout the paper we assume that the disk lies in the $z=0$ plane and work in a cylindrical
coordinate system ($r, \varphi, z$).
Starting from the full 3D hydrodynamic equations the vertically averaged equations,
describing the disk evolution in the $r-\varphi$ plane, are obtained
by integrating over the vertical direction.
The continuity equation then reads
\begin{equation}
	\int \frac{\partial \rho}{\partial t} \,dz + \int \nabla \cdot ( \rho \vec{v} ) \,dz = 0\,,
\end{equation}
where $\vec{v}$ and  $\rho$ are the 3D velocity and density, respectively.
This is typically written as
\begin{equation}
	\frac{\partial \Sigma}{\partial t}  +  \nabla \cdot ( \Sigma \vec{u})  = 0\,,
\end{equation}
where
\begin{equation}
	\Sigma  = \int \rho \,dz
\end{equation}
is the surface density of the disk and
\begin{equation}
	\vec{u}  =  \frac{1}{\Sigma} \int ( \rho \vec{v} ) \,dz
\end{equation}
the vertically averaged 2D velocity in the disk's plane.
The integrals extend over all $z$ ranging from $- \infty$ to $+ \infty$.
The integrated momentum equation, considering only pressure and gravitational forces,
then reads
\begin{equation}
	\label{eq:momentum}
	\Sigma \frac{d\vec{u}}{dt}  = - \nabla P - \int \rho \nabla \Psi \,dz \,, 
\end{equation}
with the vertically integrated pressure $P = \int p \,dz$, where $p$ is the 3D pressure.
From Eq.~(\ref{eq:momentum}) it is obvious that the change in the velocities
is determined by the specific force (or acceleration) that is given by the ratio
of force density (the integral in Eq.~\ref{eq:momentum}) and surface density $\Sigma$, i.e.
\begin{equation}
	\label{eq:specific}
	f =  - \frac{\int \rho \nabla \Psi \,dz}{\Sigma}\,.
\end{equation}
 
We here deal with the correct treatment of the last term in Eq.~(\ref{eq:momentum}), which describes the
averaging of the potential that can be given, for example, by a central star, an embedded planet, or self-gravity
\begin{equation}
	\Psi = \Psi_\mathrm{*} + \Psi_\mathrm{p} + \Psi_\mathrm{sg}\,.
\end{equation}
From the derivation of the momentum equation (\ref{eq:momentum}) it is clear that not the potential in the
midplane matters, but a suitable averaging of the gravitational force.
In the following we will analyze this averaging using density stratifications in the vertical isothermal case.
Even though accretions disks are known not to be isothermal, we nevertheless prefer at this stage
to use this assumption because it is still a commonly used approach that avoids solving a more complex energy
equation. It leads to a Gaussian density profile that is also in more general cases a reasonable first-order
approximation. Additionally, the irradiation of a central light source tends to make disks
more isothermal in the vertical direction. More realistic structures will be treated in future work.

\section{The locally isothermal accretion disk model}
\label{sec:iso-model}

The vertical structure of accretion disks is determined through the hydrostatic balance
in the direction perpendicular to the disk's midplane.
In this section we present first an idealized situation, where the disk is non-self-gravitating
and its structure is not influenced by an embedded planet. This allows for analytic evaluation of integrals.
Then we will consider more general cases.
Using the thin-disk approximation and a gravitational potential generated by a
point mass $M_*$ in the center, i.e. $\Psi_* = - G M_* /(r^2 + z^2)^{1/2}$, the vertical hydrostatic
equation for long distances $r$ from the star can be written as
\begin{equation}
	\label{eq:rho-iso}
	\frac{1}{\rho} \frac{\partial p}{\partial z}
	= - \frac{G M_* z}{\left(r^2 + z^2\right)^\frac{3}{2}}
	\approx - \frac{G M_* z}{r^3}
	\equiv - \Omega_\mathrm{K}^2 z \, .
\end{equation}
To obtain the vertical stratification for the density, $\rho$, from this equation requires
an equation of state $p=p(\rho, T)$ and 
a detailed thermal balance, by considering, for example, internal (viscous) heating,
stellar irradiation and radiative transport through the disk. To simplify matters,
often the so-called locally isothermal approximation is applied for the (numerical) study of
embedded planets in disks or for self-gravitating disks.
Using this approach, a costly solution of the energy equation is
avoided by specifying a priori a fixed radial temperature distribution $T(r)$. 
At each radius, however, the temperature is assumed to be
isothermal in the vertical direction. 
Using 
\begin{equation}
	p = \rho c_\mathrm{s}^2\,,
\end{equation}
where the (isothermal) sound speed $c_\mathrm{s}$ is now independent of $z$,
we obtain for the vertical disk structure a Gauss profile
\begin{equation}
	\label{eq:rhogauss}
	\rho_{\rm G} (r, \varphi, z) = \rho_0 (r, \varphi)  \, e^{ - \frac{1}{2} \frac{z^2}{H^2} } \,,
\end{equation}
where the vertical height $H$ is defined as
\begin{equation}
\label{eq:height}
	H = \frac{c_\mathrm{s}}{\Omega_\mathrm{K}} \,,
\end{equation}
and $\rho_0 (r, \varphi)$ is the density in the midplane of the disk.
For the vertically integrated surface density $\Sigma$ one obtains in this case
\begin{equation}
\label{eq:sigma}
	\Sigma = \int \rho_0 e^{ - \frac{1}{2} \frac{z^2}{H^2}}  \,dz = \sqrt{2 \pi} H \rho_0 \,.
\end{equation}
In 2D, $r-\varphi$, simulations of disks
this surface density $\Sigma (r, \varphi)$ is one of the evolved quantities.

\section{The potential of an embedded planet}
\label{sect:planet}
To illustrate the necessity of potential- (or force-) smoothing we consider in this section embedded planets in
locally isothermal disks that are non-self-gravitating. First we will look at an unperturbed disk whose structure
is determined by the distance from the host star and is given by Eq.~(\ref{eq:rhogauss}).
Then we include the effects of an embedded planet on the disk structure.

\subsection{Unperturbed disk}
We consider the potential of a point like planet with mass $ M_\mathrm{p} $ embedded in a 3D locally isothermal
disk with a fixed Gaussian density profile, $\rho_{\rm G}$. The planetary potential is given by
\begin{equation}
	\label{eq:planetpot}
	\Psi_\mathrm{p}(\vec{r})
	= - \frac{G M_\mathrm{p}} {\left| \vec{r} - \vec{r}_\mathrm{p} \right|}
	=  - \frac{G M_\mathrm{p}} {\left( s^2 + z^2 \right)^\frac{1}{2}} \,,
\end{equation}
where $\vec{r}$ is the vector pointing to the location in the disk and
$\vec{r}_{\rm p}$ is the vector pointing to the planet.
The vector $\vec{s}$ pointing from the disk element to the projected position of the planet, i.e. in the $ z= 0$ plane, is denoted by
\begin{equation}
	\label{eq:svector}
	\vec{s} = (\vec{r}_\mathrm{p} - \vec{r}) - \langle \vec{r}_\mathrm{p} - \vec{r} , \vec{e}_z \rangle \, \vec{e}_z \,,
\end{equation}
where $ \vec{e}_z $ is the unit vector in $ z $ direction 
and $\langle \cdot , \cdot \rangle$ denotes the scalar product.

\begin{figure}
	\centering
	\includegraphics[width=\columnwidth]{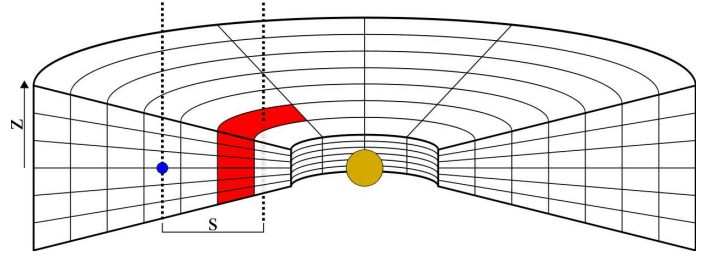} 
	\caption{
		\label{fig:disk_model_planet}
		Geometry of protoplanetary disk with embedded planet. We calculated the gravitational force from the planet (blue)
		on a vertical slice of the disk (red). For that purpose the gradient of the potential has to
		be vertically integrated along the dashed line that goes through the cell center. To obtain the total force
		exerted on the shaded disk segment, this value has to be multiplied by its area in the $r-\varphi$ plane.
	}
\end{figure}

The force acting on each disk element is then calculated from the gradient
of the potential. Since it acts along the vector $\vec{s}$, we can write
for the modulus of the vertically averaged force density
\begin{equation}
	\label{eq:force_s}
	F_\mathrm{p}(s)
	=  - \int \rho \frac{\partial \Psi_\mathrm{p}}{\partial s} \,dz
	=  - G M_\mathrm{p} s \int \frac{\rho}{(s^2 + z^2)^\frac{3}{2}} \,dz \,.
\end{equation}
The force density (with units force per area) in Eq.~(\ref{eq:force_s}) 
has to be evaluated at the centers of all individual grid cells, as illustrated in Fig.~\ref{fig:disk_model_planet}.
From there, either the total force of a disk element can be calculated by multiplying $F_\mathrm{p}(s)$ with the
disk element's area ($r \Delta r \Delta \varphi$), or one computes the specific force $f_\mathrm{p} = F_\mathrm{p} / \Sigma$,
which can be used directly to update the velocities, see Eq.~(\ref{eq:specific}). 
For the density we write
\begin{equation}
	\label{eq:densvert}
	\rho_{\rm G} (r, \varphi, z) = \rho_0 (r,\varphi) \cdot \rho_z (z^2/H^2) \,,
\end{equation}
where we have assumed that the vertical dependence of $\rho$ is a function of
the quantity $z^2/H^2$, as stated in Eq.~(\ref{eq:rhogauss}).
Substituting $y = z/s$ in Eq.~(\ref{eq:force_s}) and using a vertical stratification according to
Eq.~(\ref{eq:densvert}), we find 
\begin{equation}
	\label{eq:F_p}
	F_\mathrm{p} = - \frac{G M_\mathrm{p} \rho_0}{s} \cdot 2 I_\mathrm{p}(s)\,. 
\end{equation}
The dimensionless function $I_\mathrm{p}(s)$ is defined through an integral over half the disk
\begin{equation}
	\label{eq:i_p}
	I_\mathrm{p}(s) = \int_0^{\infty} \frac{\rho_z(c^2 y^2)}{{(1 + y^2)}^\frac{3}{2}}  \,dy\,,
\end{equation}
where the normalized vertical distance $y$ and the quantity $c$ are given by
\begin{equation}
	\label{eq:yc}
	y = \frac{z}{s}
	\quad  \mbox{and}  \quad 
	c  =  \frac{s}{H} \,.
\end{equation}

For the standard Gaussian vertical profile, i.e. $\rho_z(c^2 y^2) = \exp \left(-\frac{1}{2} c^2 y^2\right) $,
the integral $I_\mathrm{p}(s)$ can be expressed as
\begin{equation}
\label{eq:I_bessel}
	I_\mathrm{p}(s) = \frac{1}{4} c^2 \exp \left( \frac{c^2}{4} \right) \left[ K_1\left( \frac{c^2}{4} \right) - K_0\left( \frac{c^2}{4} \right) \right] \,,
\end{equation}
where $ K_n(x) $ are the modified Bessel functions of the second kind.
For illustrating purposes we present evaluations of the force correction function $I_\mathrm{p}$ for
simpler polynomial density stratifications in Appendix \ref{app:b}.

In 2D numerical simulations of disks the above averaging procedure is typically not performed.
Instead, an equivalent 2D potential is used in the momentum equation such that
\begin{equation}
	\label{eq:2dpot}
	\frac{d \vec{u}}{dt} = - \nabla \Psi^\mathrm{2D}_\mathrm{p}\,.
\end{equation}
We point out that a 2D potential with the property of equation Eq.~(\ref{eq:2dpot}) cannot be the result of
an averaging procedure in general as in Eq.~(\ref{eq:momentum}) because for realistic densities 
\begin{equation}
	\label{eq:pot2d_ave}
	\nabla \Psi^\mathrm{2D}_\mathrm{p} \,  \neq  \, \frac{\int \rho \nabla \Psi \,dz}{\Sigma} \,.
\end{equation}
For $\Psi^\mathrm{2D}_\mathrm{p}$ a simple smoothed version is often used, which reads
\begin{equation}
	\label{eq:planet_2dpot}
	\Psi^\mathrm{2D}_\mathrm{p} = - \frac{G M_\mathrm{p}}{\left(s^2 + \epsilon_\mathrm{p}^2\right)^\frac{1}{2}} \,,
\end{equation}
where $\epsilon_\mathrm{p}$ is the smoothing length to the otherwise
point mass potential, introduced to avoid numerical problems at the location of
the planet. We will refer to this functional form of $\Psi$ as the $\epsilon$-potential,
although it is sometimes named Plummer-potential as well.
The force acting on each disk element is then calculated from the gradient of the potential.

A finite $\epsilon_\mathrm{p}$ regularises the potential and guarantees stable numerical evolution.
Additionally, it serves to account for the vertical stratification of the disk.
Comparing torques acting on a planet in 2D and 3D simulations it has been
suggested for the Lindblad torques that $\epsilon_\mathrm{p}$ should be on the order of the vertical disk height,
specifically $0.7 H$, see \citet{2002A&A...387..605M}.
He pointed out, however, that for the corotation torques a lower value may be appropriate.
Hence, often a value of $\epsilon = 0.6 H$ is chosen.
Here, we calculated the correct smoothing by a vertical average assuming an isothermal, vertically stratified disk.
This will lead to a distance-dependent smoothing.

\begin{figure}
	\centering
	\includegraphics[width=\columnwidth]{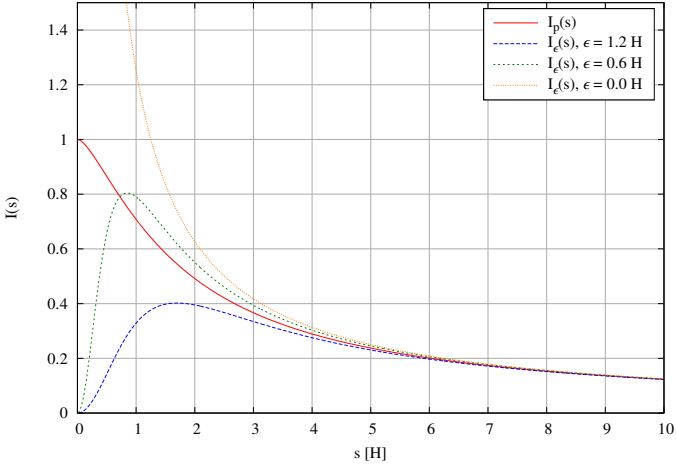} 
	\caption{
		\label{fig:int_planet}
		Force correction function $I_\mathrm{p}(s)$ (solid line) resulting from an integration over the
		vertical structure of the disk, see Eqs.~(\ref{eq:F_p}, \ref{eq:I_bessel}). Additional curves indicate
		the corresponding function for the 2D potential (equivalent to Eq.~(\ref{eq:2dpot-correspond})) using different
		values of the smoothing parameter $\epsilon_\mathrm{p}$.
	}
\end{figure}

In Fig.~\ref{fig:int_planet} we compare the force correction obtained from the vertically
averaging procedure and the 2D smoothed $\epsilon$-potential. Specifically, we plot the
function $I_\mathrm{p}(s)$ for the Gaussian density profile together with the
corresponding function for the 2D potential,
which reads 
\begin{equation}
	\label{eq:2dpot-correspond}
	I_\epsilon(s) = \frac{s^2}{\left(s^2+\epsilon^2\right)^\frac{3}{2}} \frac{\sqrt{2 \pi} H}{2} \,.
\end{equation}
Because we are mainly interested in distances $s $ up to a few $H$, we assume, that the disk height $ H$ does not change with radius.
The unsmoothed $\epsilon_\mathrm{p}=0$ potential diverges as $1/s$ for short distances from the planet, leading to
a $1/s^2$ force as expected for a point mass. Since the value of $I_{\rm p} (s)$ remains finite for $s \rightarrow 0$,
we see that a vertically extended disk reduces the divergence of the force to $1/s$. 
This is a clear indication that in 2D simulations the potential has to be smoothed for physical reasons alone, and that
the assumption of a point mass potential will greatly overestimate the forces.
As expected, the $\epsilon$-potential strongly reduces the force for $s \rightarrow 0$ and always yields regular
conditions at the location of the planet.
Additionally, it appears that the value $\epsilon_\mathrm{p} = 0.6 H$ overestimates the potential slightly
for $s \gtrsim H$ (green curve in Fig.~\ref{fig:int_planet}).

\begin{figure}[htb]
	\centering
	\includegraphics[width=\columnwidth]{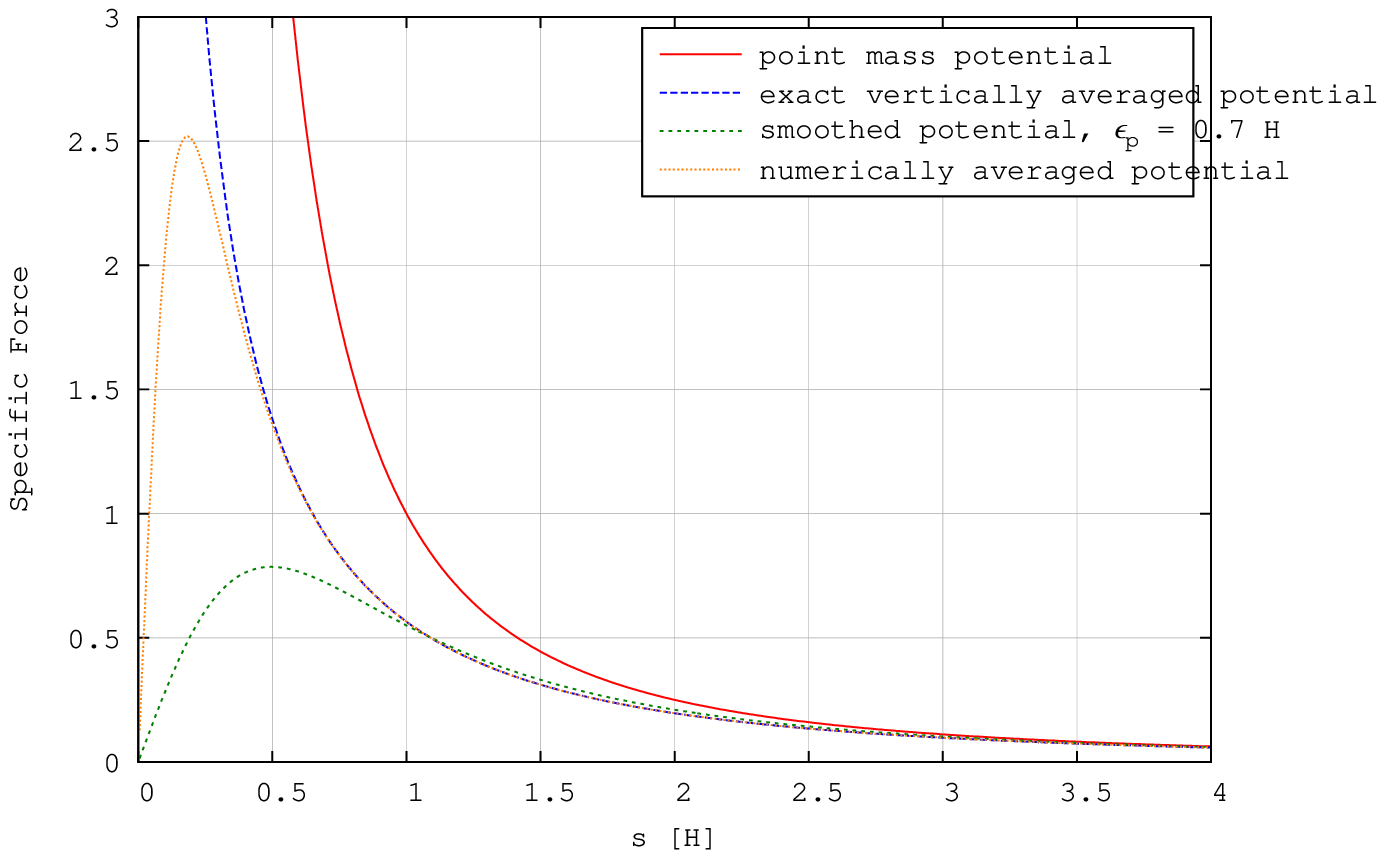}
	\caption{
		\label{fig:F_planet}
		Specific gravitational force $F_\mathrm{p}/\Sigma$ as a function of distance from the planet.
		Different approximations are shown: 
		a) an ideal point mass (solid-red) that falls off as $1/r^2$;
		b) the exact averaged force according to Eqs.~(\ref{eq:fs_p}) and (\ref{eq:I_bessel});
		c) the force according to the smoothed potential of Eq.~(\ref{eq:planet_2dpot}) using
		here $\epsilon_{\rm p} = 0.7 H$; and
		d) a numerically integrated averaged force using 10 grid points in the vertical direction
		and a maximum $z_{\rm max}=5 H$. 
		The force is normalized such that $G M_{\rm p}=1$.
	}
\end{figure}

Using the Gaussian isothermal density distribution, $\rho_{\rm G}$, and the surface density $\Sigma$ from Eq.~(\ref{eq:sigma}),
the force density of the planet acting on the disk can be written as
\begin{equation}
	\label{eq:fs_p}
	F^{\rm G}_\mathrm{p} = - \frac{G M_\mathrm{p} \Sigma}{s}  \frac{1}{H}  \, \sqrt{ \frac{2}{\pi} } \, I_\mathrm{p}(s)\,.
\end{equation}
This expression could in principle be used directly in numerical simulations of planet-disk interaction.
However, even though $ I_\mathrm{p}(s) $ can be solved in terms of Bessel-functions, in computational hydrodynamics this evaluation is not very efficient,
because it has to be calculated once per timestep at each grid point. 
A possibility is to solve the integral $I_\mathrm{p}(s)$ in Eq.~(\ref{eq:i_p}) directly numerically using a limited number of
vertical grid cells. In this case the integral is converted into a sum, where the exponential can be pre-calculated and
tabulated at the corresponding nodes. For our purposes we found that only 10 vertical grid points give an adequate solution (see below).
This is shown in Fig.~\ref{fig:F_planet} where we plot the specific force (acceleration) 
exerted by a planet on a disk element that is a distance $s$ away. The force for a point mass falls off as $1/s^2$ while for small radii the exact vertically averaged force 
shows a $1/s$ behavior. Two approximations are displayed as well: The curve for the $\epsilon$- potential
given in Eq.~(\ref{eq:planet_2dpot}), where we used a constant $\epsilon_\mathrm{p}=0.7 H$. The numerically averaged curve refers to a numerical
integration of $I_\mathrm{p}(s)$ using 10 grid points in the $z$-range $ [0, 5 H]$. Note that because of the finite discretization 
no additional smoothing is required. Increasing the number of grid points increases the agreement with the exact averaged force
even more for shorter distances $s$. The scaling of the distance with $1/H$ and the normalization of the force
make the plot independent of the used vertical thickness of the disk.
While the $\epsilon$-potential agrees well for $s \gtrsim H$ with the exact averaged force,
our numerical approximation agrees to much shorter distances.
In Appendix \ref{app:b} we show that simplified density distributions can lead to an equally good agreement.
However, the advantage of the described vertical numerical integration procedure lies in its speed and in the fact that it leads directly
to a regularized force for very short $s$. Also, it is easily generalizable, as will be shown in the next section.

\begin{figure}
	\centering
	\includegraphics[width=\columnwidth]{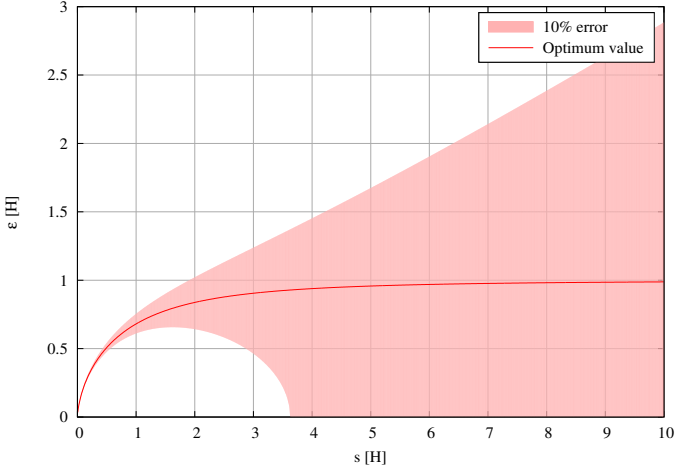} 
	\caption{
		\label{fig:epsilon_planet}
	Optimum value of $ \epsilon_\mathrm{p} $ in the smoothing potential of Eq.~(\ref{eq:planet_2dpot}) as
       a function distance $ s $ from the planet. The shaded area illustrates
		which values of $ \epsilon_\mathrm{p} $ result in a force error of less than $10\, \%$ with respect
     to the exact averaged value.
	}
\end{figure}

Because we can calculate $ I_\mathrm{p}(s) $ for the Gaussian profile numerically to any required accuracy,
we can estimate the optimum $ \epsilon_\mathrm{p}(s) $ value for the smoothed potential in Eq.~(\ref{eq:planet_2dpot})
for each point to obtain agreement with the exact averaged force.
In Fig.~\ref{fig:epsilon_planet} we display the correct $ \epsilon_\mathrm{p}(s) $ value against the distance. The
range of $ \epsilon_\mathrm{p} $ in which the smoothed potential produces an error of less than $10\,\%$
in the force compared to the correct value is illustrated by the shaded region.
Obviously for very short distances it is important to use the correct $ \epsilon_\mathrm{p} $ 
and for long distances the exact value of $ \epsilon_\mathrm{p} $ does not play an important role.
Through a Taylor expansion for $ s \to \infty $ of the denominator in Eq.~(\ref{eq:i_p}) of $ I_\mathrm{p}(s)$ it can be shown that 
\begin{equation}
	\lim_{s \to \infty} \epsilon_\mathrm{p}(s) = H \,,
\end{equation}
which can be expected because the disk has a vertical extent on the order $ H $.

\subsection{Taking the planet into account}
\label{subsec:planet-rho}
In the previous section we have analyzed the forces acting on the planet considering an
unperturbed disk with a given scale height as determined by the star, see Eq.~(\ref{eq:height}). However, an embedded planet
changes the disk structure, which will lead, for example, to a reduced thickness in the vicinity of the planet.
This plays an important role for the torques acting on the planet.
To estimate this effect we start from the vertical hydrostatic equation (\ref{eq:rho-iso}) taking an additional planet
into account,
\begin{equation}
	\label{eq:rho-iso-planet}
	\frac{1}{\rho} \frac{\partial p}{\partial z}
	= - \frac{G M_* z}{\left(r^2 + z^2\right)^\frac{3}{2}}
	- \frac{G M_{\rm p} z}{\left(s^2 + z^2\right)^\frac{3}{2}} \,,
\end{equation}
where $s$ is again the distance from the planet.
For a vertically isothermal disk this can be integrated
\begin{equation}
	\rho_{\rm p} = \rho_0  \exp\left\{- \frac{1}{2} \frac{G M_*}{c_\mathrm{s}^2 r^3} z^2  
		+ \frac{G M_{\rm p}}{c_\mathrm{s}^2 s} \left[ \frac{s}{(s^2 + z^2)^{1/2}} - 1 \right] \right\}\,,
\end{equation}
where we assumed for the stellar contribution $z \ll r$, as before.
Using the previous vertical thickness $H$ (Eq.~(\ref{eq:height})) and the mass ratio $q=M_{\rm p}/M_*$, this can be written as
\begin{equation}
	\label{eq:exact_rho}
	\rho_{\rm p}  = \rho_0  \exp\left\{- \frac{1}{2} \frac{z^2}{H^2}   
		+  q \frac{r^3}{H^2 s} \left[ \frac{s}{(s^2 + z^2)^{1/2}} - 1 \right] \right\} \,.
\end{equation}
Because $s$ is on the same order as $z$ in the neighborhood of the planet, this equation cannot be
simplified further. To still obtain an estimate of the expected effects, we {\it approximate} the
vertical density stratification by 
\begin{equation}
	\label{eq:simple_rho}
	\rho_{\rm p}^{\rm a} = \rho_0 \exp\left\{- \left( \frac{1}{2} \frac{z^2}{H^2} + \frac{|z|}{H_\mathrm{p}} \right) \right\}\,,
\end{equation}
where we define the reduced scale height near the planet 
\begin{equation}
	\label{eq:H_p}
	H_\mathrm{p} = \frac{4 s^2 H^2}{q r^3}\,.
\end{equation}
This approximation for the vertical behavior of $\rho_\mathrm{p}$ leads to the correct limits for the integrated surface density
$\Sigma$ in the limits for very short and long $s$ and is a reasonable approximation in between.
From the definition of $H_\mathrm{p}$ in Eq.~(\ref{eq:H_p}) we find that the distance $s_{\rm t}$ where the two
scale heights are equal, i.e. $H_\mathrm{p} = H$, is given by 
\begin{equation}
	\label{eq:s_t}
	\frac{s_{\rm t}}{r}  =  \frac{1}{2} \, \left(\frac{q}{h}\right)^{1/2}\,, 
\end{equation}
where $h$ is the relative scale height, $h=H/r$, of the disk.   
The location $s_{\rm t}$ denotes approximately a transitional distance from the planet. For $s \lesssim s_{\rm t}$ the standard
approximation of a Gaussian density distribution with the scale height $H$ is no longer valid, and the
influence of the planet dominates.

\begin{figure}
	\centering
	\includegraphics[width=\columnwidth]{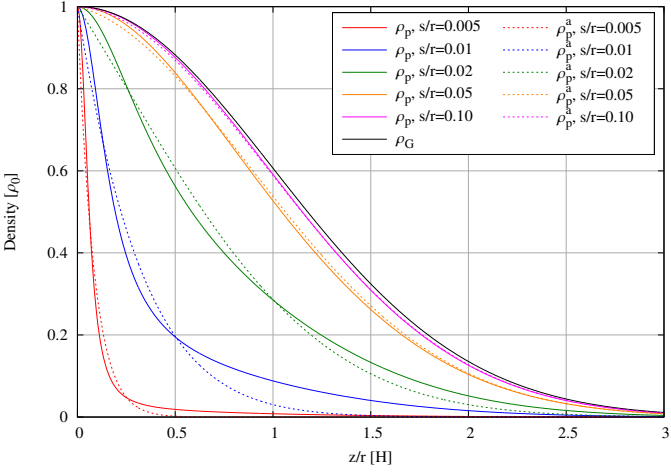} 
	\caption{
		\label{fig:rhot}
		Exact (see Eq.~\ref{eq:exact_rho}) and approximate (see Eq.~\ref{eq:simple_rho}) vertical density profiles for a
	disk with aspect ratio $ h=0.05 $ for short distance $ s $ to a planet with mass ratio $ q = 6 \times 10^{-5} $. 
	The exact profiles are shown by the solid lines and the corresponding approximate ones in the same color but with
	dashed lines.
	For comparison the unperturbed Gaussian profile (see Eq.~\ref{eq:rhogauss}) is shown in black for this $H/r$.
}
\end{figure}

Fig.~\ref{fig:rhot} shows the exact (Eq.~\ref{eq:exact_rho}) and the approximate (\ref{eq:simple_rho}) vertical density
profiles for different distances to the planet and the comparison with the unperturbed Gaussian profile.
For very short distances to the planet the effective height of the disk is much lower than in the unperturbed case.
Obviously, $\rho_{\rm p}^{\rm a}$ is only an approximation to $\rho_{\rm p}$, but it captures the change in thickness of the disk,
as induced by the planet, very well.
We will show below that using $\rho_{\rm p}^{\rm a}$ in the force calculations yields a very good approximation to the
exact case. 

\begin{figure}[htb]
	\centering
	\includegraphics[width=\columnwidth]{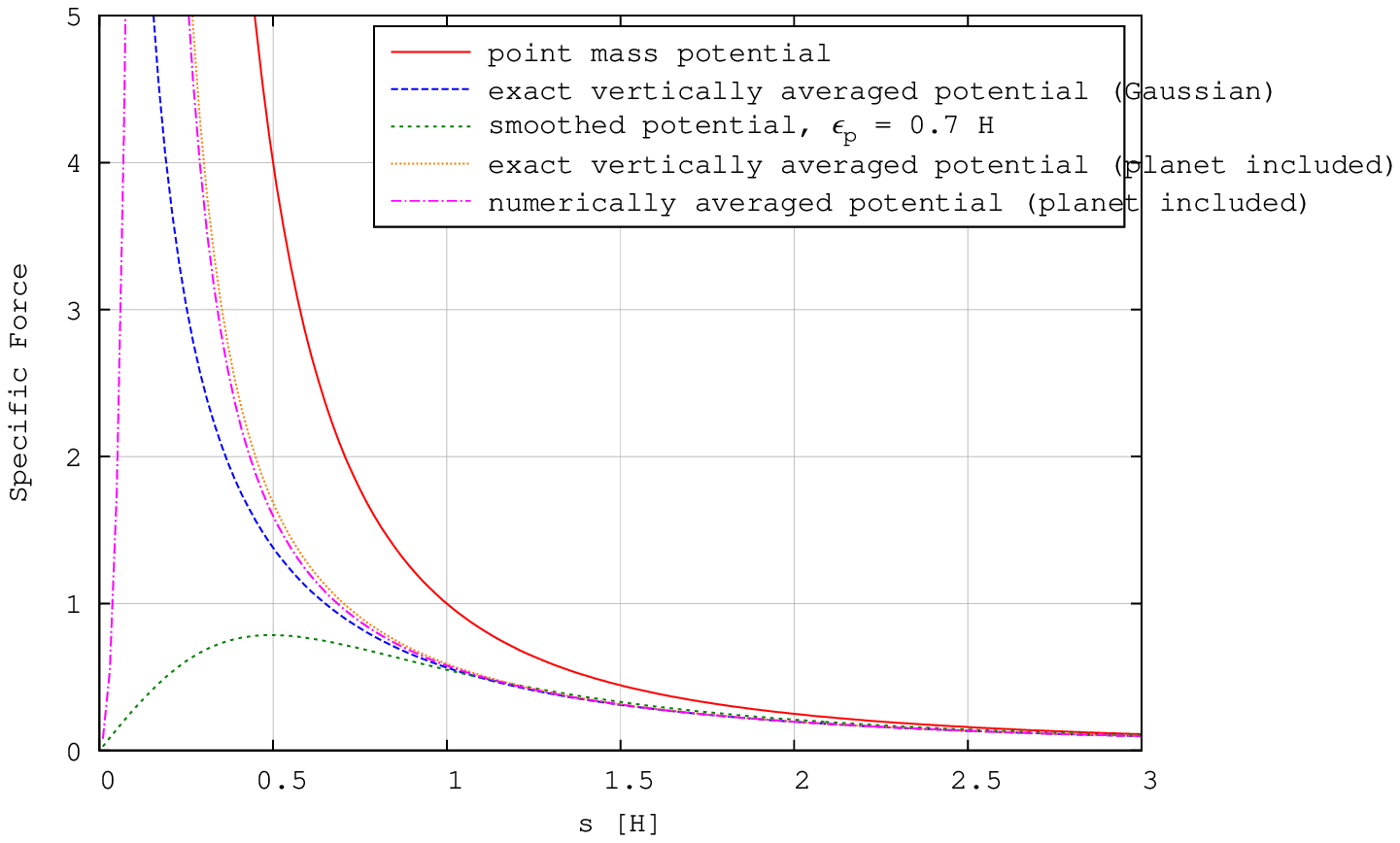}
	\caption{
		\label{fig:force-pc}
		Specific force $F_\mathrm{p}/\Sigma$ for different approximations. 
		The first three curves are identical to Fig.~\ref{fig:F_planet} and are shown
		for comparison. The 'exact' potential using $\rho_\mathrm{p}$ is shown in yellow and the approximates solution, using the simplified expression for the
		density $\rho_\mathrm{p}^\mathrm{a}$ and only 10 vertical grid points, is shown in pink. 
		A tapering function near the center was used for the latter to regularize the force.
	}
\end{figure}

To test how this new density-scaling influences the previous isothermal results for the planet's gravity,
we calculated a corrected vertically averaged force using the modified density $\rho_\mathrm{p}$ according to 
\begin{equation}
	\label{eq:force_pc}
	F_\mathrm{p} (s)
	=  - 2 G M_\mathrm{p} s \int_0^{z_{\rm max}} \frac{\rho_\mathrm{p}}{(s^2 + z^2)^\frac{3}{2}} \,dz \,.
\end{equation}
This integral has to be calculated numerically using an approximate $z_{\rm max}$. 
Depending on the model to be calculated (either 'exact' or approximate force),
we substituted either $\rho_\mathrm{p}$ from Eq.~(\ref{eq:exact_rho}) or the approximate
$\rho_\mathrm{p}^\mathrm{a}$ from Eq.~(\ref{eq:simple_rho}). Since the presence of the
planet alters the vertical height of the disk, $z_{\rm max}$ depends on the distance $s$ from the planet.
To estimate  $z_{\rm max}$  we first define an effective vertical scale height
\begin{equation}
\label{eq:heff}
	H_{\rm eff} =  \left( \frac{1}{H^2}  + \frac{1}{H_{\rm p}^2}  \right)^{-1/2}\,,
\end{equation}
which is an interpolation of the value at short and long distances from the planet.
We choose to take $z_{\rm max} = 6 H_{\rm eff}$ in the 'exact' numerical evaluation where we use
the density $\rho_\mathrm{p}$ and 1000 grid cells. In contrast, we apply $z_{\rm max} = 3 H_{\rm eff}$  in the approximate numerical
evaluation using $\rho_\mathrm{p}^\mathrm{a}$ and only 10 vertical grid cells, see also Appendix \ref{app:a}.

In Fig.~\ref{fig:force-pc} we display the results for the vertically integrated specific force $F_\mathrm{p}/\Sigma$ in various approximations.
The first three curves (red, blue, green) correspond to those shown in Fig.~\ref{fig:F_planet} as an illustration.
The 'exact' potential using $\rho_\mathrm{p}$ is shown in yellow and the approximate solution using $\rho_\mathrm{p}^\mathrm{a}$ in pink.
The presence of the planet reduces the scale height of the disk, which leads to an enhancement of the force above the Gaussian,
i.e., it diverges as $s^{-2}$ for short distances.
To regularize the approximate force, we added a tapering function that reduces it to zero in the vicinity of the planet 
(pink dashed-dotted curve) such that it can be used in hydrodynamic simulations, see Appendix \ref{app:a}.
Clearly the range of applicability for short $s$ is much improved over the simple $\epsilon$-potential. 
In contrast to the Gaussian approximation that uses the fixed thickness $H$,
the force correction (with respect to the pure point mass potential) depends now on the mass of the planet as well,
which enters through $H_\mathrm{p}$.

\subsection{Numerical simulations of planet-disk interactions}
\label{subsect:numerics-planet}

\begin{figure}
	\centering
	\includegraphics[width=\columnwidth]{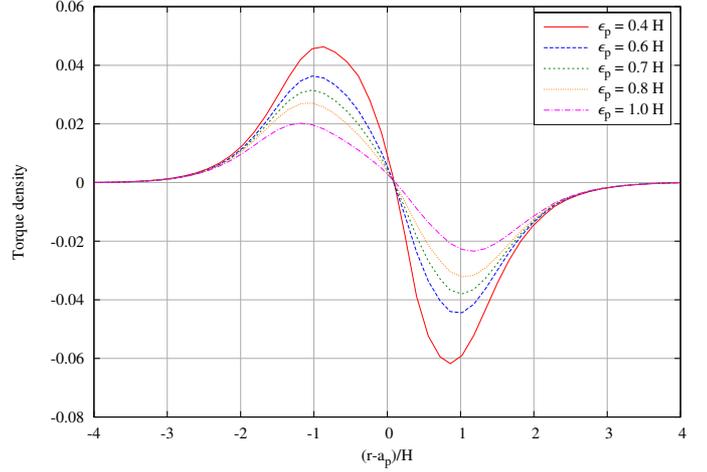}
	\caption{
		\label{fig:gam2}
		Specific radial torque density in units of $(d\Gamma/dm)_0$ (see Eq.~\ref{eq:gamm0})
		for embedded planet models using the $\epsilon$ potential with various $\epsilon_\mathrm{p}$.
		The simulations use a given $H=0.05$, and the $x$-axis refers to the radial coordinate where $a_{\rm p}$ is the
		semi-major axis of the planet. The planet to star mass-ratio is $q=6 \times 10^{-5}$. Other parameters of the
		simulations are stated in Sect.~\ref{subsect:numerics-planet}
	}
\end{figure}

\begin{figure}
	\centering
	\includegraphics[width=\columnwidth]{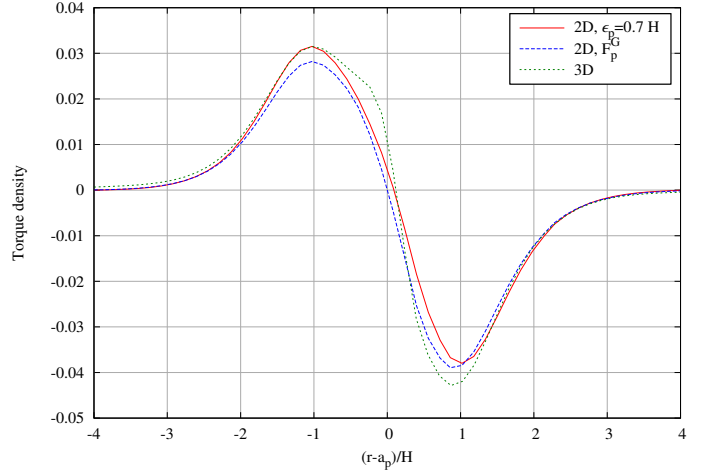}
	\caption{
		\label{fig:gamrcn}
		Specific radial torque density in units of $(d\Gamma/dm)_0$ for embedded planet models for 
		2D models using $\epsilon$-potential (red curve) and
		a numerically integrated force $F_\mathrm{p}^\mathrm{G}$ (see Eq.~\ref{eq:fs_p}) that takes into account a Gaussian 
		vertical density distribution (blue curve).
		The green curve refers to a full 3D model using the same physical model parameter as the 2D models. The
		3D result is adapted from \citet{2009A&A...506..971K} where the setup and numerics is described in more detail.
	}
\end{figure}

To test the formulation of the force, we performed numerical simulations
of embedded planets in two and three dimensions where we use an isothermal equation
of state. For this purpose we solved the 2D and 3D hydrodynamic equations for a viscous gas. 
We used a setup very similar to that of \citet{2008A&A...487L...9K} and \citet{2009A&A...506..971K}.
The planet with a mass ratios $M_{\rm p}/M_* = 6 \times 10^{-5}$ is embedded at $r=1$ in the disk with radial
extent of $0.4$ -- $2.5$. The disk is locally isothermal such that the aspect ratio $H/r$ is constant. This implies
a radial temperature gradient of $T(r) \propto r^{-1}$, while in the vertical direction $T$ is constant. 
As a consequence, the unperturbed vertical density structure (in the 3D simulations) is Gaussian along the $z$-axis. 
The surface density falls off with radius as $\Sigma \propto r^{-1/2}$.
We used a constant kinematic viscosity coefficient of $\nu = 10^{-5}$ in dimensionless units. 
This setup is such that without the planet the disk is exactly in equilibrium and would not evolve with time.
At the radial boundaries damping boundary conditions were applied \citep{2006MNRAS.370..529D}.
In our simulations we used $H/r=0.05$ and evolved the disk after the planet's insertion for about 100 orbits.
To test our improved treatment of the force, we varied the planet mass and the scale height of the disk in comparison models.
For the standard parameter, $h=0.05$ and $q=6 \times 10^{-5}$, we found for the transition distance $s_{\rm t} \approx 0.35 H$.

The embedded planet disturbs the disk and torques are exerted on it by the disk through gravitational back-reaction.
These might lead to a change in the planet's orbital parameter.
The strength of these torques will depend on the applied smoothing of the gravitational force.
To illustrate the effect, it is convenient to study the radial torque distribution per unit disk mass
$d \Gamma(r)/dm $, which we define here, following \citet{2010ApJ...724..730D}, such that the total torque $\Gamma_{\rm tot}$
is given as 
\begin{equation}
	\Gamma_{\rm tot} = 2 \pi \int \frac{d \Gamma}{d m} (r) \, \Sigma(r) \, r dr \,.
\end{equation}
In other words, $d \Gamma(r)$ is the torque
exerted by a disk annulus of width $dr$ located at the radius $r$ and having the mass $dm$.
As $d \Gamma(r)/dm$ scales with the mass ratio squared and as $(H/r)^{-4}$, we rescale our results accordingly in units of
\begin{equation}
	\label{eq:gamm0}
	\left( \frac{d \Gamma}{d m}\right)_0 =  \Omega_\mathrm{p}^2(a_\mathrm{p})\,a_\mathrm{p}^2  q^2 \left(\frac{H}{a_\mathrm{p}}\right)^{-4}\,,
\end{equation}
where the index $p$ denotes that the quantities are evaluated at the planet's position, with the semi-major axis $a_\mathrm{p}$.
%% Note that \citet{2010ApJ...724..730D} used a $(H/a)^{-4}$ scaling.

In Fig.~\ref{fig:gam2} we plot $d \Gamma(r)/dm$ obtained from 2D simulations using the $\epsilon$-potential for the
gravity of the planet and the standard fixed scale height $H$ for the disk.
The torque in this and the similar following plots are scaled to  $(d \Gamma(r)/dm)_0$ as given in Eq.~(\ref{eq:gamm0}).
Results for five values of the smoothing length are presented.
Obviously, the value of $\epsilon_\mathrm{p}$ has great impact on the amplitude of the torque density, and making the correct choice
is important.
We point out that the differences in the torque density also influence the total torques that determine the
important migration rate. Because $\Gamma_{\rm tot}$ consists of positive and negative contributions of similar magnitude,
even small errors in $d \Gamma(r)/dm$ can lead to large errors in $\Gamma_{\rm tot}$. For the range of $\epsilon$ displayed in
Fig.~\ref{fig:gam2} we find a variation of $\Gamma_{\rm tot}$ larger than about a factor of 4. The best agreement in this
case of $H/r=0.05$ with 3D results is obtained for $\epsilon$ in between $0.6-0.7H$. 

In  Fig.~\ref{fig:gamrcn} we compare two different force treatments for 2D simulations to a full 3D run.
Results for the $\epsilon_\mathrm{p}=0.7H$ potential are given by the red curve, and the blue curve corresponds to 
the vertically averaged force $F_\mathrm{p}^\mathrm{G}$ according to Eq.~(\ref{eq:fs_p}), which assumes a Gaussian vertical density profile.
In magnitude the $\epsilon_\mathrm{p}=0.7H$ potential represents the vertically averaged results reasonably well,
but it behaves differently close to the planet.
The additional green curve corresponds to a full 3D (locally) isothermal simulation as presented in \citet{2009A&A...506..971K}
(their Fig.~10, purple curve).
The 3D runs use an identical physical setup and a more realistic cubic-potential with a smoothing of $r_{\rm sm}=0.5$. 
Both 2D results are on the same order as the 3D run, but show small deviations that can lead to larger variations in the total torque,
end hence the migration rate,
because the positive and negative contributions to the radial torque density are of similar magnitude.

\begin{figure}
	\centering
	\includegraphics[width=\columnwidth]{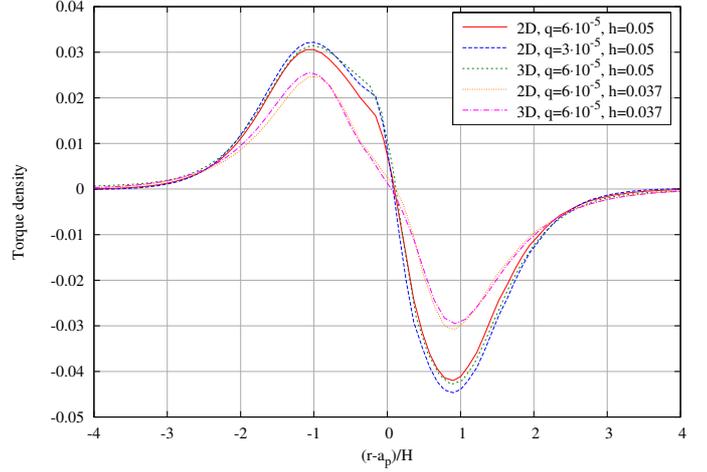}
	\caption{
		\label{fig:gam2-pc}
		Specific radial torque density in units of $(d\Gamma/dm)_0$ for 2D and 3D embedded planet models using
		different mass ratio and disk temperature.
		The first three curves represent models with the same disk temperature $H/r = 0.05$. The red curve corresponds to
		our standard model, the blue curve has a reduced planet mass $q=3 \times 10^{-5}$, and the third (green curve) 
		corresponds to a full 3D run. The next two curves (yellow and purple) refer to models with $H/r = 0.037$.
		The 2D runs used our approximate density distribution $\rho_\mathrm{p}^\mathrm{a}$ in evaluating the gravitational force.
		The 3D results were adapted from \citet{2009A&A...506..971K} where the setup and numerics is described in more detail.
	}
\end{figure}

\begin{figure}
	\centering
	\includegraphics[width=\columnwidth]{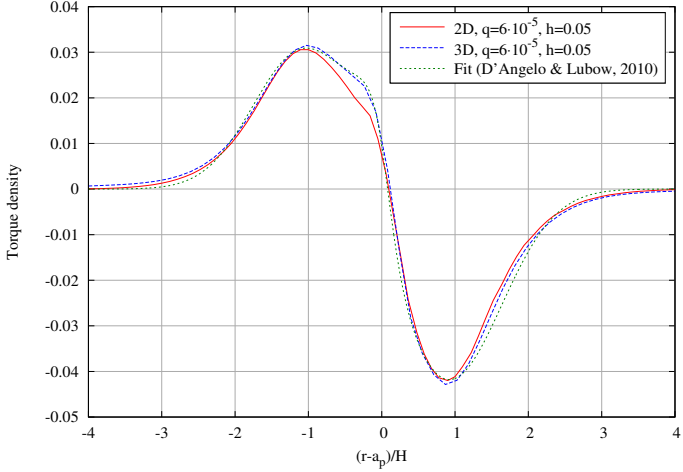}
	\caption{
		\label{fig:gam3-pc}
		Specific radial torque density in units of $(d\Gamma/dm)_0$ for 2D and 3D embedded planet models using
		different mass ratio and disk temperature.
		The first two curves are identical to those in the previous figure, and the third is a fit
		presented by \citet{2010ApJ...724..730D} corresponding to the model with $T(r) \propto r^{-1}$ 
		and $\Sigma(r) \propto r^{-1/2}$.
        }
\end{figure}

To test the applicability of our new procedure for treating the gravity in 2D simulations, we performed runs with different
mass ratio and temperature in the disk. The results of 2D and 3D simulations are shown in Fig.~\ref{fig:gam2-pc}, where the
torque density $d\Gamma(r)/dm$ is plotted. The first set of simulations refers to $H/r = 0.05$, where we compared the standard
model in 2D and 3D to a run with half the planet mass. The next two curves show results of 2D and 3D simulations
for $H/r = 0.037$. 
Despite the indicated differences in the parameter, all models used the same physical setup as described above.
The 2D runs used our approximate density distribution $\rho_\mathrm{p}^\mathrm{a}$ in evaluating the gravitational force.
The 3D results are adapted from \citet{2009A&A...506..971K} where the setup and numerics is described in more detail.
Firstly, all five curves show very similar overall behavior, confirming the scaling with  $(d\Gamma(r)/dm)_0$. 
The reduced amplitude of the $H/r=0.037$ models is due to the onset of gap formation.
Secondly, the agreement of the 2D and 3D runs is very good indeed. For example, 
upon varying the scale height, the change in shape of the curves is identical in 2D and 3D runs (yellow and purple) curve.
We point out that the value of $\epsilon$ to obtain the best agreement of the total torque $\Gamma_{\rm tot}$ 
in 2D and 3D simulations may depend on the value of $H$, because of the influence of the planet. Hence, it is always advisable
to perform the simulations using the vertical integration of the force.

To additionally validate our simulations we compare in Fig.~\ref{fig:gam3-pc} our 2D and 3D results obtained for the standard model
to an analytic fit by \citet{2010ApJ...724..730D} for the same disk parameter.
Although the fit has been developed for a smaller planet mass, the agreement of the two 3D results is excellent. This
is interesting because our planet mass of 20 $M_{\rm Earth}$ is already in the range where non-linear effects should set in.
The 2D torque shows the same amplitude, but small differences are visible just inside the planet.
In isothermal disks the flow close to the planet is not in exact hydrostatic equilibrium anymore. However, full 3D high-resolution
isothermal simulations \citep{2003ApJ...586..540D}  have shown that vertical velocities are only large inside the Roche region of the planet.
This explains the good agreement of the 2D approximation with the full 3D case.

\section{The potential of a self-gravitating disk}
\label{sec:disk-potential}

\begin{figure}
	\centering
	\includegraphics[width=\columnwidth]{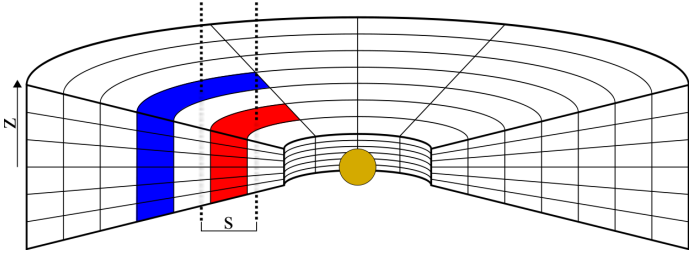} 
	\caption{
		\label{fig:disk_model_disk}
		Geometry of a protoplanetary disk for calculations with self-gravity. 
		We calculate the gravitational force exerted by a vertical slice of the disk (blue)
		on another vertical slice of the disk (red), separated by a distance $s$.
		Two vertical integrations have to be performed along the dashed lines that go through the cell centers.
		To obtain the total force between the two segments, this value has to be multiplied by the corresponding areas,
		see also Eq.~(\ref{eq:force_sg}).
	}
\end{figure}

Now we turn to full self-gravitating disks where a smoothing of the gravitational potential is required as well
to account for the finite thickness of the disk.
The potential at a point $\vec{r}$ generated by the whole self-gravitating disk is given by
\begin{equation}
	\label{eq:pot_sg}
	\Psi_\mathrm{sg}(\vec{r})
	= - \int_{\rm Disk}  \frac{G \rho(\vec{r}')}{\left| \vec{r} - \vec{r}' \right|} \,d\vec{r}' \,.
\end{equation}
The smoothing required to obtain the potential in the midplane has been analyzed for this situation
by \citet{2009A&A...507..573H}. However, if one is interested in problems of fragmentation, it is more convenient to analyze the force between to 
individual elements (segments) of the disk. Let us consider the force between two such disk segments that are a separated by
the distance $\vec{s}$, see Fig.~\ref{fig:disk_model_disk}. 
The potential at the location $\vec{r}$ generated by a disk element located at $\vec{r}'$ which is a
projected distance $s$ away is given by
\begin{equation}
	\Psi_\mathrm{sg}(\vec{r}) = - \iiint \frac{G \rho(r', \varphi', z')}{\left( s^2 + (z-z')^2 \right)^\frac{1}{2}} \,dz'\,dA' \,,
\end{equation}
where $dA'$ is the surface element of the disk in the $r-\varphi$-plane at the point $\vec{r}'$.
The vector $\vec{s}$ is defined in analogy to Eq.~(\ref{eq:svector}) and illustrated in Fig.~\ref{fig:disk_model_disk}.

The force at the location $\vec{r}$ generated by this vertically extended disk element is calculated from
the gradient of the potential. The vertically averaged force density can then be written in analogy to Eq.~(\ref{eq:force_s}) as
\begin{align}
\label{eq:force_sg}
	F_\mathrm{sg}(s)
	&= -\int \rho(r,\varphi,z) \frac{\partial \Psi_\mathrm{sg}}{\partial s} \,dz  \nonumber  \\ 
	&= -Gs \iiiint \frac{\rho(r',\varphi',z')\,\rho(r,\varphi,z)}{\left( s^2 + (z-z')^2 \right)^\frac{3}{2}} \,dz'\,dA'\,dz \,.
\end{align}

The evaluation of this integral depends on the vertical stratification of the density at both locations $\vec{r}$ and $\vec{r}'$.
As before, we consider locally isothermal disks. For weakly self-gravitating disks the density structure is then given by the
Gaussian form in Eq.~(\ref{eq:rhogauss}). However, similar to an embedded planet the disk's self-gravity will modify the
vertical profile. Following our previous treatment of the embedded planet,
we first analyzed the smoothing required for a disk that has an unperturbed Gauss profile and will then allow for modifications.

\subsection{Unperturbed disk}
\label{subsec:unperturbed}
For a locally isothermal disk and with Eq.~(\ref{eq:densvert}) we obtain
\begin{equation}
	\label{eq:F_d}
	F_\mathrm{sg}(s) = -  G \rho_0(r, \varphi) \, \iint \rho_0(r', \varphi') \,dA' \cdot 2 I_\mathrm{sg}(s) \,,
\end{equation}
where we defined the function $ I_\mathrm{sg}(s) $ by
\begin{equation}
	\label{eq:i_d}
	I_\mathrm{sg}(s) = \frac{1}{2} \int_{-\infty}^\infty \int^\infty_{-\infty} \frac{\rho_z(c^2 y^2)\,\rho_z(c^2 y'^2) }{\left( 1 + (y-y')^2 \right)^\frac{3}{2}} \,dy' \,dy \,,
\end{equation}
where the normalized vertical distance $ y $ and the quantity $ c $ are given by Eq.~(\ref{eq:yc})
again assuming a constant $H$.
This integral cannot be calculated directly for the standard Gaussian profile, and we will evaluate it numerically.

In 2D numerical simulations usually a simple smoothed potential is used instead of calculating the correct averaging.
This $\epsilon_\mathrm{sg}$-potential reads as
\begin{equation}
	\label{eq:disk_2dpot}
	\Psi^\mathrm{2D}_\mathrm{sg}(s)  = - \iint \frac{G \Sigma(\vec{r}')}{\left(s^2 + \epsilon_\mathrm{sg}^2\right)^\frac{1}{2}} \,dA' \,,
\end{equation}
where $ \epsilon_\mathrm{sg} $ is the smoothing length.
The force acting on each disk element is then calculated from the gradient of $\Psi^\mathrm{2D}_\mathrm{sg}$.

\begin{figure}
	\centering
	\includegraphics[width=\columnwidth]{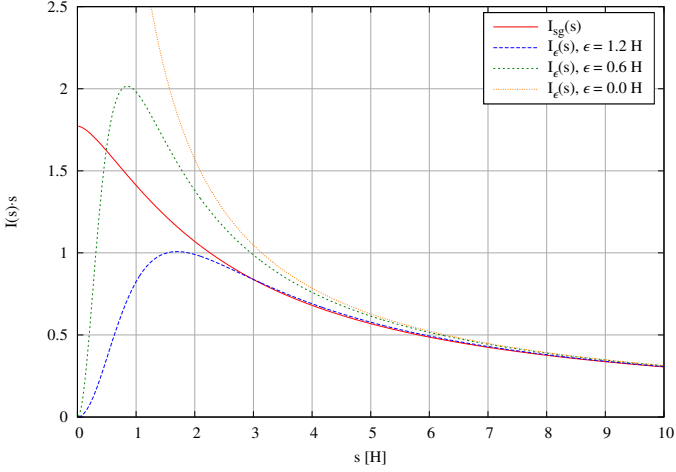} 
	\caption{
		\label{fig:int_disk}
		Force correction function $I_\mathrm{sg}(s)$ multiplied by $ s $ resulting from an integration over the
		vertical structure of the disk, see Eqs.~(\ref{eq:F_d},\ref{eq:i_d}). Additional curves indicate
		the corresponding function for the 2D potential (Eq.~\ref{eq:2dpot-correspond}) using different
		values of the smoothing parameter $\epsilon_\mathrm{sg}$.
	}
\end{figure}

In Fig.~\ref{fig:int_disk} we compare the force correction obtained from the vertically averaging procedure and
the 2D smoothed potential for a disk height that does not change with radius.
Because $ I_\mathrm{sg}(s) $ diverges for short distances, we multiply it by $ s $. Because
we are interested in local effects with distances $ s $ up to a few $H,$ we assumed for the plot
that the disk height $ H $ is constant.
Comparing this result to the corresponding force correction function for the planet in Fig.~\ref{fig:int_planet}, it is obvious
that for the self-gravitating case a larger smoothing is required for the $\epsilon$-potential than in the planet case.
Here, a value of $\epsilon = 1.2$ seems
to be a good choice. Lower values, already $\epsilon=0.6$,  considerably overestimate the force.
We attribute this larger required smoothing with the double vertical averaging that has to be performed in this case.

\begin{figure}
	\centering
	\includegraphics[width=\columnwidth]{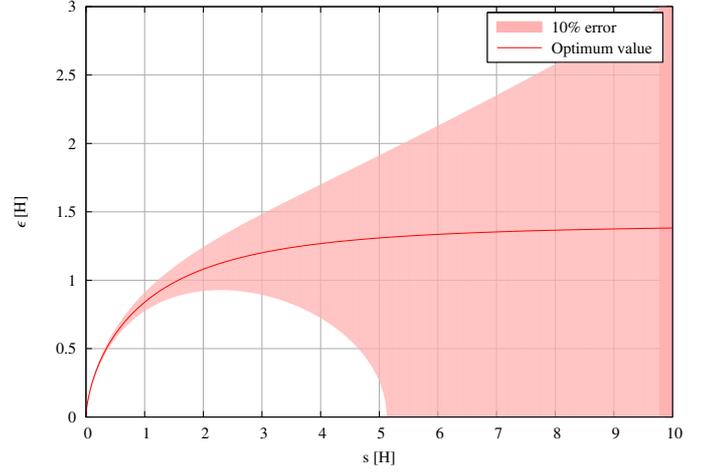} 
	\caption{
		\label{fig:epsilon_disk}
		Optimum value of $ \epsilon_\mathrm{sg} $ in the smoothing potential of Eq.~(\ref{eq:disk_2dpot}) as a function of distance $ s $ using a Gaussian vertical stratification. The colored area illustrates
		which values of $ \epsilon_\mathrm{sg} $ result in an error of the force correction value of less than $ 10\,\%$.
	}
\end{figure}

From the numerically calculated $ I_\mathrm{sg}(s) $ for the Gaussian profile we can calculate the best $ \epsilon_\mathrm{sg}(s)$ value for the smoothing potential in Eq.~(\ref{eq:disk_2dpot}) at each distance $s$ to obtain the right force correction value.
In Fig.~\ref{fig:epsilon_disk} we display the optimum $ \epsilon_\mathrm{sg}(s) $ value versus distance.
The range of $ \epsilon_\mathrm{sg} $ over which the smoothing potential produces an error of less than $10\,\%$ is shaded.
Cleary for short distances it is crucial to use the correct value of $ \epsilon_\mathrm{sg}$,
whereas for long distances the influence of $ \epsilon_\mathrm{sg} $ becomes negligible.
Because we now have to account twice for the vertical extent of the disk 
here compared to the planet case, we obtain a higher limiting value of
\begin{equation}
	\lim_{s \to \infty} \epsilon_\mathrm{sg}(s) = \sqrt{2} H \,.
\end{equation}
This value was obtained through numerical calculation up to the sixth significant digit for $ s = 1000 H $, and we verified
it through a Taylor expansion of the denominator in Eq.~(\ref{eq:disk_2dpot}).
As before, it is interesting that the required optimum smoothing remains finite, even for very long distances.
This result agrees with \citet{2009A&A...507..573H}.

\subsection{Taking the disk into account}
\begin{figure}
	\centering
	\includegraphics[width=\columnwidth]{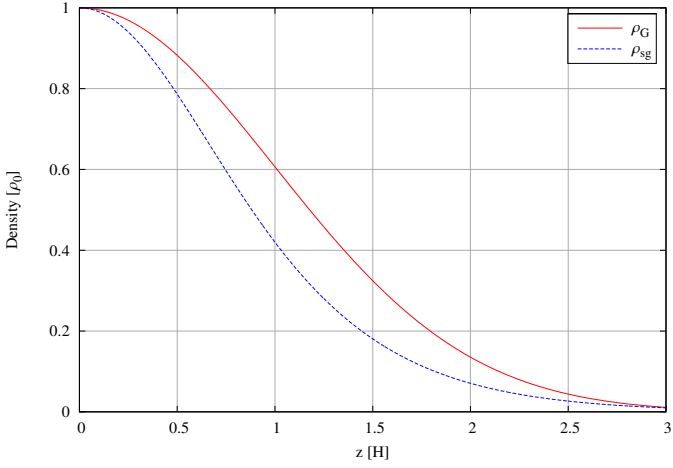} 
	\caption{
		\label{fig:profiles2}
		Vertical density profiles of the disk with for a self-gravitating disk (see Eq.~\ref{eq:rho_sg}) with $ Q = 1 $. For comparison the unperturbed Gaussian profile (see Eq.~\ref{eq:rhogauss}) is shown in red.
}
\end{figure}

Now we consider the correction for a disk where self-gravity has modified the vertical structure. For that purpose, we considered 
the case of a pure self-gravitating disk, neglecting the gravitational potential of the central mass. Assuming a large disk with a
slowly varying surface density $ \Sigma(r)$, then 
the derivate with respect to  $ z $ of the disk potential given in Eq.~(\ref{eq:pot_sg}) simplifies \citep{1963MNRAS.126..553M} to
\begin{equation}
	\frac{\partial \Psi_\mathrm{sg}}{\partial z} = 2 \pi G \Sigma(r) \,,
\end{equation}
and consequently the hydrostatic equation (\ref{eq:rho-iso}) changes to
\begin{equation}
	\frac{1}{\rho} \frac{\partial p}{\partial z} = - 2 \pi G \Sigma(r)\,.
\end{equation}
For a vertically isothermal disk this can be integrated \citep{1942ApJ....95..329S} to
\begin{equation}
	\label{eq:rho_sg}
	\rho_\mathrm{sg}(r, \varphi, z) = \rho_0(r, \varphi) \, \cosh^{-2} \left(\frac{z}{H_\mathrm{sg}}\right) \,,
\end{equation}
where the vertical scale height $ H_\mathrm{sg} $ is defined by
\begin{equation}
	H_\mathrm{sg} = \frac{c_\mathrm{s}^2}{\pi G \Sigma} = \frac{c_\mathrm{s} \Omega_\mathrm{K}}{\pi G \Sigma} H \cong Q H \, ,
\end{equation}
with the Toomre parameter $Q$ \citep{1964ApJ...139.1217T}. Fig.~\ref{fig:profiles2} shows the changed vertical density profiles
in the self-gravitating case compared to the unperturbed Gaussian for $ H = H_\mathrm{sg} $ or $ Q = 1 $, respectively.
The self-gravitating profile is steeper and therefore, for equal surface density,
the mass is consequently located nearer to the midplane of the disk.
This is expected because the vertical component of the disk's gravitational potential is,
with $ 2 \pi G \Sigma(r) \sim GM/r^2 $, by a factor of about $ r/H $ 
larger than in Eq.~(\ref{eq:rho-iso}), and so the mass is more concentrated toward the midplane. 

\begin{figure}
	\centering
	\includegraphics[width=\columnwidth]{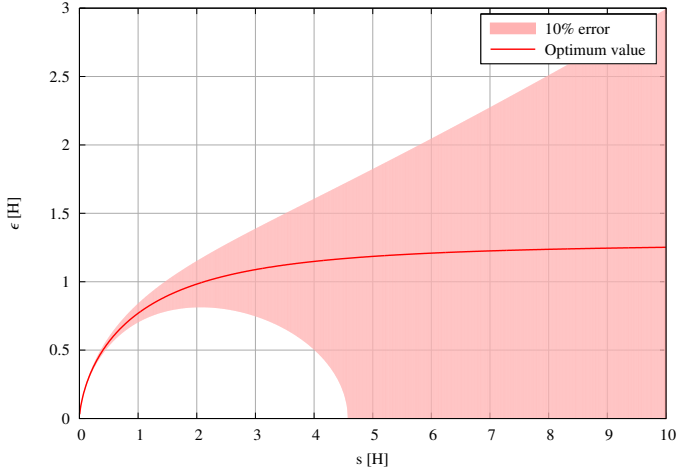} 
	\caption{
		\label{fig:epsilon_disk_sg}
	Optimum value of $ \epsilon_\mathrm{sg} $ in the smoothing potential of Eq.~(\ref{eq:disk_2dpot}) versus distance $ s $ using a vertical stratification caused by self-gravity with $ Q = 1$. The colored area illustrates
		which values of $ \epsilon_\mathrm{sg} $ result in an error of the force correction value of less than $ 10\,\%$.
	}
\end{figure}

Now we can calculate the force correction function $ I_\mathrm{sg}(s) $ of equation (\ref{eq:i_d}) with the self-gravitating vertical density profile $ \rho_\mathrm{sg} $.
In analogy to the unperturbed case, this can be used to calculate an optimum $ \epsilon_\mathrm{sg} $ for the $ \epsilon$-potential.
In Fig.~\ref{fig:epsilon_disk_sg} we display the correct $ \epsilon_\mathrm{sg}(s) $ value against distance.
The optimum $ \epsilon_\mathrm{sg} $ for this value of $Q$ is always lower than our previous unperturbed Gaussian case.
For the limit $ s \to \infty $ we find a value about $ 10\,\%$ lower.
This is in consistency with the lower effective vertical scale height, because more mass is located near the midplane.

In most self-gravitating disks the vertical structure will be affect by both the central mass object and the self-gravity of the disk. Then it is to be expected that the correct $ \epsilon_\mathrm{sg} $ is a value between
the two extreme cases. 
The combined situation where self-gravity and the central mass both contribute has been considered by 
\citet{2007NCimR..30..293L}. For clarity, we treat the two cases separately here.

\begin{figure}
	\centering
	\includegraphics[width=\linewidth]{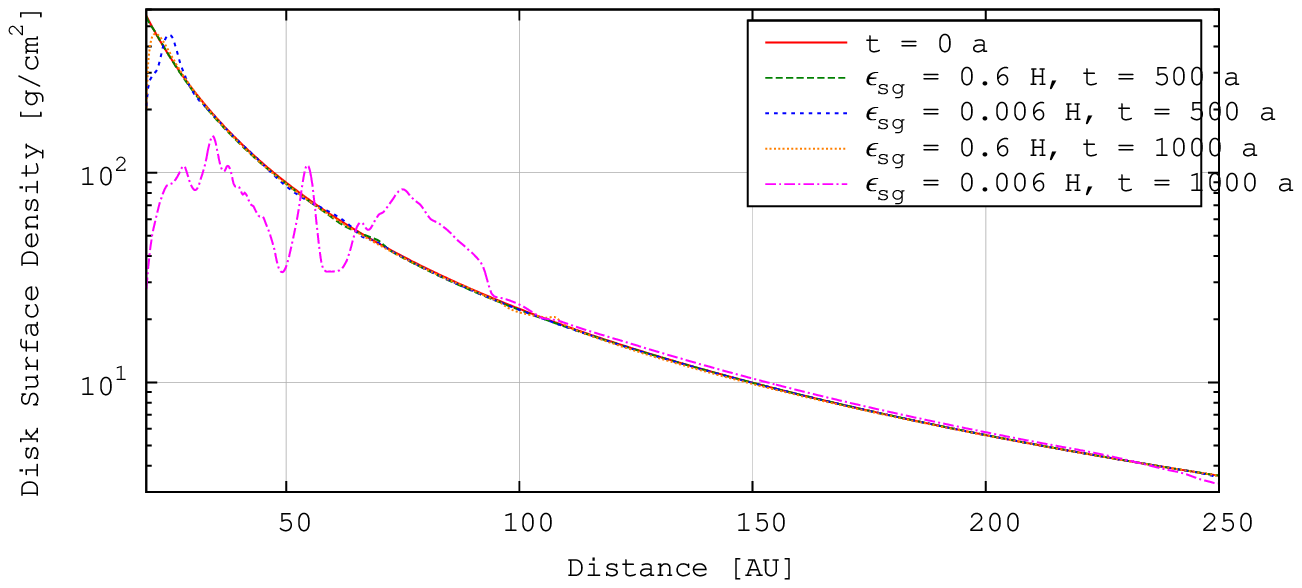} 
	\includegraphics[width=\linewidth]{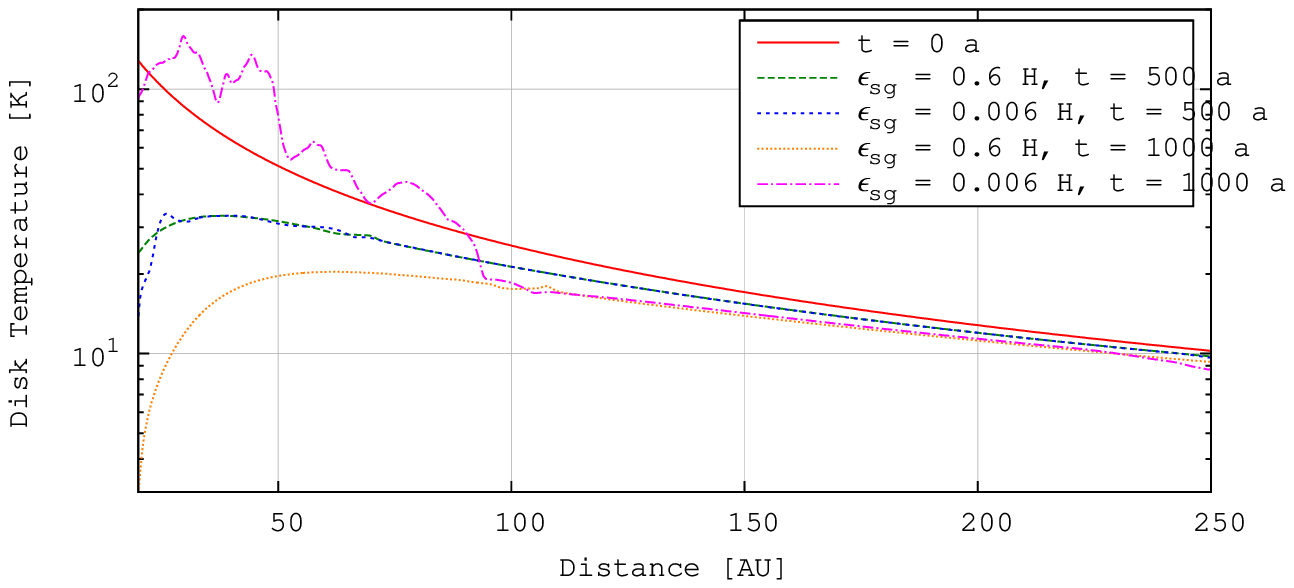} 
	\includegraphics[width=\linewidth]{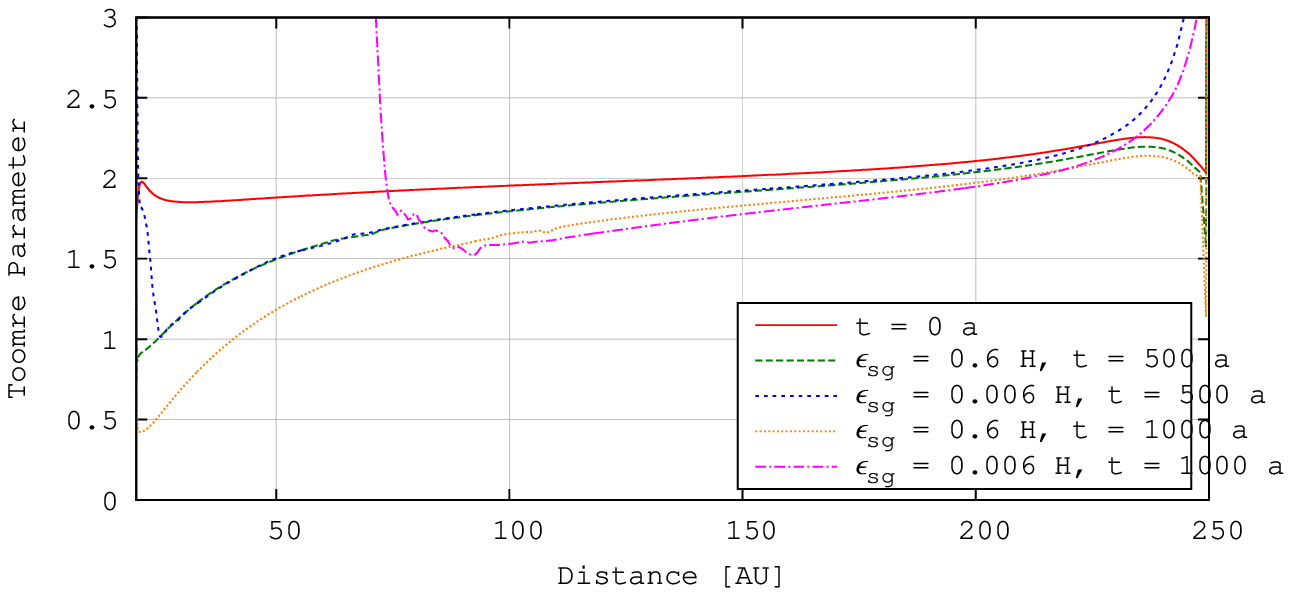} 
	\caption{
		\label{fig:2d_disk_model_profiles}
		Radial dependency of surface density (\textit{top}), midplane temperature (\textit{middle}) and Toomre parameter (\textit{bottom}) of the two disk models with $ \epsilon_\mathrm{sg} = 0.6 H $ and 
		$\epsilon_\mathrm{sg} = 0.006 H $ for different timestamps. At $ t = 0\,$a both simulations match because they have identical start conditions. In the first $ 500 $ years they behave very similar but then
		their further evolution diverges drastically.
	}
\end{figure}

\subsection{Numerical simulations of self-gravitating disks}

\begin{table}
	\caption{
		\label{tab:parameters}
		Parameters of the disk model. The top entry refers to host star. The following give the
		disk properties, and then the initial disk setup and the computational parameter are given.
	}
	\centering
	\renewcommand\arraystretch{1.2}
	\begin{tabular}{ll}
		\hline 
		Star mass ($ M_\mathrm{primary} $) & $ 1.0\,M_{\sun} $ \\
		\hline
		Disk mass ($ M_\mathrm{disk} $) & $ 0.4\,M_{\sun} $ \\
		Adiabatic index ($ \gamma $) & $ 5/3 $ \\
		Mean-molecular weight ($ \mu $) & $ 2.4 $ \\
		$\beta$-Cooling ($ \beta $) & $ 20 $ \\
		\hline
		Initial density profile ($ \Sigma $) & $ \propto r^{-2} $ \\
		Initial temperature profile ($ T $) & $ \propto r^{-1} $ \\
		Initial disk aspect ratio ($ H/r $) & $ 0.1 $ \\
		\hline
		Grid ($N_r \times N_\varphi$) & $ 512 \times 1536 $ \\
		Computational domain ($ R_\mathrm{min} $ -- $ R_\mathrm{max} $) & $20$ -- $250\,\mathrm{AU} $ \\
		\hline
	\end{tabular}
\end{table}

To demonstrate the influence of $ \epsilon_\mathrm{sg} $ in numerical simulations, 
we analyzed the effects of different values of $ \epsilon_\mathrm{sg} $ on the fragmentation conclusions of a self-gravitating disk.
For that purpose, we adopted a disk model from \citet{2011MNRAS.416.1971B}.
Table~\ref{tab:parameters} shows all
important disk parameters. We simulated the disk twice using the \texttt{ADSG} version of the \texttt{FARGO} code \citep{2008ApJ...678..483B, 2000A&AS..141..165M} for
$ 5000 $ years with values of $ 0.6 H $ and $ 0.006 H $ for $ \epsilon_\mathrm{sg} $. Both models include the self-gravity of the disk and a simple $\beta$-cooling model,
which is defined by 
\begin{equation}
	\frac{\partial E}{\partial t} = - \frac{E \Omega}{\beta} \,,
\end{equation}
where $ E $ is the internal energy, $ \Omega $ the angular velocity and $ \beta $ a constant.
The disk must cool fast enough to be able to fragment, which is otherwise prevented by compressional heating.
\citet{2001ApJ...553..174G} showed that this is the case for $ \beta \lesssim 3 $ and \citet{2005MNRAS.364L..56R}
found later a dependency from the equation of state for $ \beta $ and suggested a $ \beta \lesssim 7 $ for $ \gamma = 5/3 $. 

\begin{figure*}
	\centering
	\includegraphics[width=\textwidth]{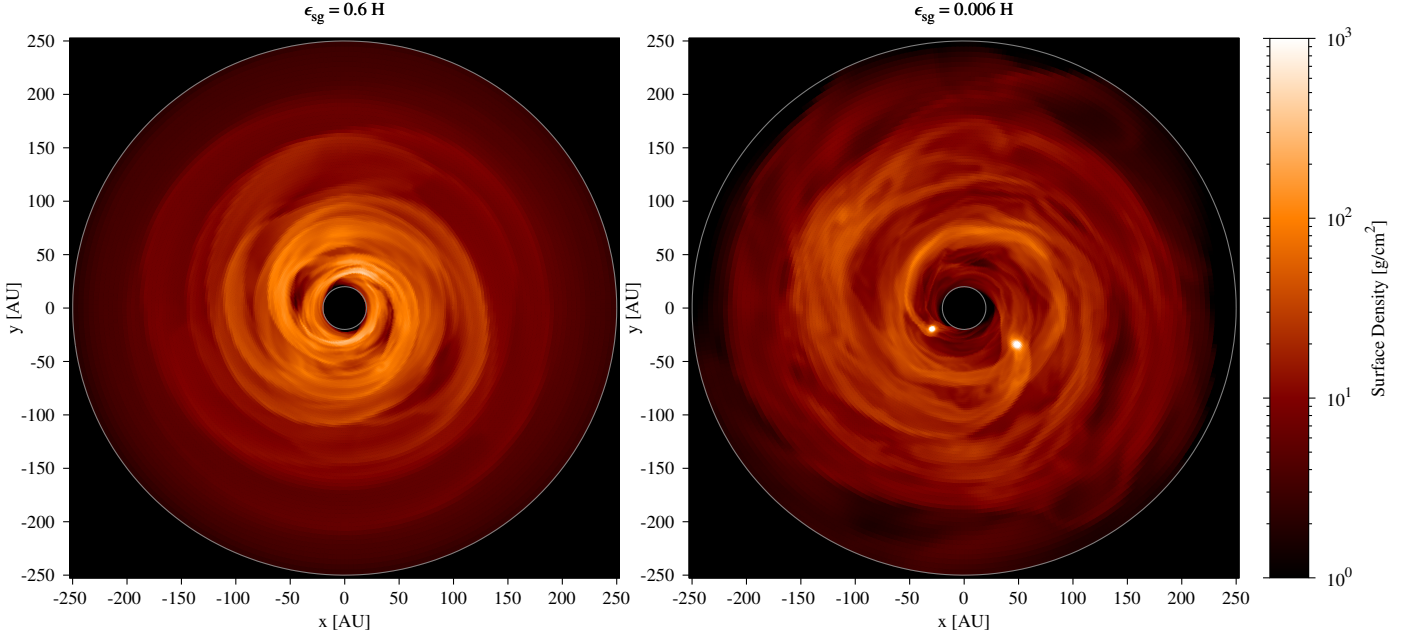} 
	\caption{
		\label{fig:density_maps}
		Surface density of two disks after $ 5000 $ years which started from the same initial condition but with different values for the smoothing parameter $ \epsilon_\mathrm{sg} $ of the
		gravitational potential. The gray circle shows the computational domain.
	}
\end{figure*}
\citet{2011MNRAS.411L...1M} pointed out that the previous results had not converged with increasing resolution and that
the critical value, $ \beta_\mathrm{crit} $, may be higher than previously thought.
They measured an $ \beta_\mathrm{crit} $ of $ \sim 18 $ for their highest resolution.
Because we use $ \beta = 20 $ in our models, we do not expect them to fragment in any case.

Another stability criterion can be described by the Toomre stability parameter $ Q $, which is defined by
\begin{equation}
	Q = \frac{\kappa c_\mathrm{s}}{\pi G \Sigma} \,,
\end{equation}
where $ c_\mathrm{s} $ is the sound speed in the disk and $ \kappa $ the epicyclic frequency.
\citet{1964ApJ...139.1217T} showed that for values of $ Q \geq 1 $ an axisymmetric disk should be stable.
As shown in Fig.~\ref{fig:2d_disk_model_profiles}, the modeled disks have an initial $ Q $ value of $ 1.85 $ -- $ 2.2 $
depending on the radial distance to the star, and thus should not be fragmenting.

In the first $ 500 $ years both disks cool down in the central parts of the disk from about $ 120 $\,K to about $ 30 $\,K, resulting in $ Q $ values of $ ~ 1 $ at the inner edge of the disk, which means
that they are not Toomre-stable anymore, but still should not fragment because the cooling constant $ \beta $ 
is higher than the critical value required for fragmentation.
After $ 500 $ years both simulations start to differ. The $ \epsilon_\mathrm{sg} = 0.006 H $ model fragments within about $ 500 $ years in the inner region of the
disk whereas, the $ \epsilon_\mathrm{sg} = 0.6 H $ model needs about $ 1000 $ years to start developing small spiral arms in the inner region.
After $ 5000 $ years two fragments have survived in the   $ \epsilon_\mathrm{sg} = 0.006 H $ disk with fragment masses of $ 16.5\,M_\mathrm{Jup} $ and $ 8.3\,M_\mathrm{Jup} $.
The $ \epsilon_\mathrm{sg} = 0.6 H $ model only shows spiral arms and no signs of fragmentation. 
Even after $ 50000 $ years we did not observe any fragments. 
Fig.~\ref{fig:density_maps} shows the surface density distribution of both models after $ 5000 $ years.

In Section~\ref{subsec:unperturbed} we suggested an $ \epsilon_\mathrm{sg} $ on the order of unity to obtain results that can compare to 3D simulations. Because the $ \epsilon_\mathrm{sg} = 0.6 H $ model
did not fragment as predicted by the stability criteria, this seems to support the validity of our estimate for $\epsilon_\mathrm{sg}$.
The very short $\epsilon_\mathrm{sg} $ in the
$ \epsilon_\mathrm{sg} = 0.006 H $ model overestimates the gravitational forces on short distances (see Fig.~\ref{fig:int_disk})
and therefore excites disk fragmentation.

\section{Summary and conclusions}
\label{sect:summary}
We analyzed the smoothing of gravity in thin 2D disk simulations for the embedded planet and self-gravitating case.
Starting from the vertically averaged hydrodynamic equations, 
we first showed  that the gravitational force has to be calculated using a density-weighted average of the 3D force, see Eq.~(\ref{eq:momentum}).
Because this depends on the density distribution, there cannot be a general equivalent 2D version of the potential
(Eq.~\ref{eq:pot2d_ave}).
To be able to explicitly calculate the averaged force, we first used a locally isothermal disk structure in which the
vertical density stratification is Gaussian. 

For the embedded planet case the resulting force can be calculated analytically. In full 2D hydrodynamic simulations we compared
the resulting torque density acting on the planet for the $\epsilon$-potential (\ref{eq:planet_2dpot})
and the 'exact' averaged force. We found that the overall
magnitude of the torque is best modeled using a smoothing of $\epsilon = 0.7 H$, while there remain significant differences to
the full 3D case, also in the total torque.
Taking the modification of the density stratification induced by the planet into account leads to a much reduced vertical thickness
of the disk in the vicinity of the planet.
We presented a simplified analytical form for the modified vertical density stratification with a planet, the details of the numerical implementation
are given in the appendix. Using this in 2D simulations leads to very good agreement of the torque density
with full 3D calculations of embedded planets on circular orbits.
By varying the disk height and the mass of the planet, we showed that the torque density scales
as expected with $(d \Gamma(r)/dm)_0$, see Eq.~(\ref{eq:gamm0}).

Because the modified density approximation that includes the planet is based on vertical hydrostatic equilibrium, it is not clear however, whether
it is valid for planets on non-circular or inclined orbits. Here, the variations occur on the orbital timescale and to establish
hydrostatic equilibrium, the thermal timescale must be on the same order.  
In the situation of multiple planets that may interact strongly, the same restrictions may apply.
Despite these restrictions, we believe that using this prescription will enhance the accuracy of 2D simulations considerably.
We expect that our procedure can be generalized to the radiative case using suitable vertical averages, but this
needs to be developed.

For the self-gravitating case we showed that the required smoothing, $\epsilon \approx H$, is even larger than in the planetary case. We attribute
this to the necessity of a double averaging over the vertical height of the disk. Owing to the complex integration, the integrals cannot
be solved analytically in this case. Taking into account self-gravity lowers the required smoothing because the vertical scale height is reduced due to the
additional gravity. In more strongly self-gravitating systems, which have a Toomre parameter $Q \approx 1$, non-axisymmetric
features may occur.
Because the standard self-gravity solvers require a smoothing that scales with radius, one has to take an approximate average
in non-axisymmetric situations.
The same applies for disks that are close to the fragmentation limit. 
As shown by our last example, the choice of smoothing may affect the conclusions on whether the disk will fragment or not.
Through detailed comparisons with full 3D simulations a suitable smoothing may be found.

\appendix

\section{Integration of the force density}
\label{app:a}
Here, we briefly outline a numerically fast and convenient method to 
vertically integrate the force for the embedded planet case.
Specifically, we plan to evaluate the force density $F_\mathrm{p}$ in Eq.~(\ref{eq:force_pc}), which reads
\begin{equation}
	\label{eq:app-force_pc}
	F_\mathrm{p} (s)
	=  - 2 G M_\mathrm{p} s \int_0^{z_{\rm max}} \frac{\rho_\mathrm{p}}{(s^2 + z^2)^\frac{3}{2}} \,dz \,.
\end{equation}
As pointed out in the text, we used for the (approximate) numerical integration a maximum $z$ of $z_{\rm max} = 3 H_{\rm eff}$,
with $H_{\rm eff}$ given by Eq.~(\ref{eq:heff}). The interval $[0,z_{\rm max}]$ is divided into $N_z$ equal intervals with
the size $\Delta z = z_{\rm max}/N_z$.
The integral in Eq.~(\ref{eq:app-force_pc}) is replaced with the following sum
\begin{equation}
	\int_0^{z_{\rm max}} \frac{\rho_\mathrm{p} \, dz}{(s^2 + z^2)^\frac{3}{2}}   \rightarrow
	\sum_{k=1}^{N_z} \, \frac{\rho_\mathrm{p}(z_k) \, \Delta z}{(s^2 + z_k^2 + r_{\rm s}^2 )^\frac{3}{2}}\,,
\end{equation}
where the $N_z$ nodes are located at $z_k = (k-1/2) \Delta z$. We introduce a small smoothing, $r_{\rm s}$, here to
keep the sum regular at short distances $s$. In the simulation presented in the paper we used $r_{\rm s} = 0.1 R_{\rm Hill}$ throughout,
which is much shorter than the unperturbed vertical height $H$. In the 2D hydrodynamic simulations we used $N_z = 10$, which makes
the method feasible numerically. 

The specific force, or acceleration ($F_\mathrm{p}/\Sigma$), is then obtained by dividing through the integrated surface density
\begin{equation}
	\Sigma =  2 \, \sum_{k=1}^{N_z} \, \rho_\mathrm{p}(z_k) \, \Delta z \,,
\end{equation}
using the same nodes. In 2D simulations $\Sigma$ is one of the evolved quantities and this relation can be used to calculate
the otherwise unknown midplane density $\rho_0$.
Because the vertical number of grid points is very small ($N_z = 10$), the method is reasonably fast and can be used in
numerical planet-disk simulations. Additionally, it only needs to be evaluated in the vicinity of the planet and could
be omitted farther out.
Nevertheless, despite the coarseness of the integration, the agreement with the 'exact' force and torque is excellent.
For numerical stability we smooth the resulting force using the following tapering function
\begin{equation}
	f_{\rm taper}(s) = \frac{1}{\exp{\left[-(s-r_{\rm t})/(0.2 r_{\rm t})\right]} + 1}
\end{equation}
with the tapering cutoff-length $r_{\rm t}$, for which we used in our simulations $r_{\rm t} = 0.2 R_{\rm Hill}$.
The purple curve in Fig.~\ref{fig:force-pc} exactly corresponds  to the procedure described here, using the stated
parameter. 

Finally, we point out that for the stellar contribution to the potential
\begin{equation}
	\label{eq:starpot}
	\Psi_\mathrm{*}(\vec{r})
	= - \frac{G M_\mathrm{*}} {\left| \vec{r} - \vec{r}_\mathrm{*} \right|}
	=  - \frac{G M_\mathrm{*}} {\left( r^2 + z^2 \right)^\frac{1}{2}}\,,
\end{equation}
a very similar vertical averaging needs to be performed. However, because the vertical profile is
always Gaussian, the nodes can be precomputed and the exponential function has only to be evaluated once for the $N_z$ nodes
per grid point and timestep.

\section{Approximate vertical density profiles}
\label{app:b}

\begin{figure}[htb]
	\centering
	\includegraphics[width=\columnwidth]{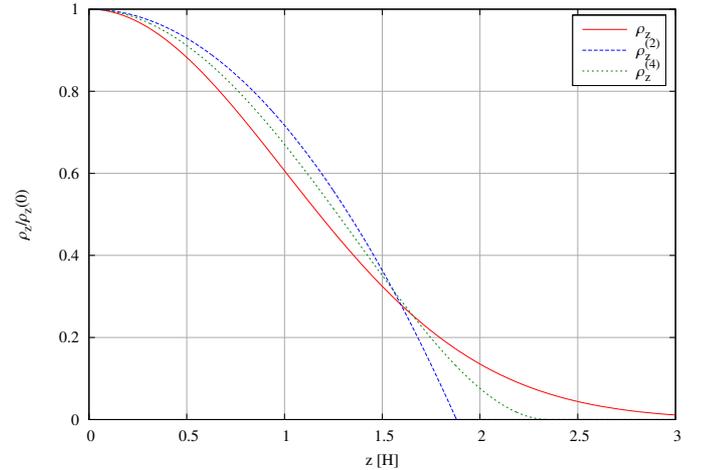}
	\caption{
		\label{fig:profiles}
		Gaussian vertical profile $\rho_z$ (solid line) compared with the parabolic vertical 
		profile $ \rho_z^{(2)} $ and fourth-order vertical profile $ \rho_z^{(4)} $.
		They all have the same area below the curves and yield the identical surface density.
	}
\end{figure}

To simplify some estimates and obtain an idea of the functional behavior of the forces, 
it is useful to study simpler vertical density stratifications. Here, we present results
for a parabolic and quartic behavior.
The corresponding density stratifications read
\begin{equation}
	\label{eq:profile_t2}
	\rho_z^{(2)} = \left[ 1 - \frac{1}{2} \left(\frac{z}{H^{(2)}} \right)^2 \right]  
\end{equation}
for the parabolic form, and 
\begin{equation}
	\label{eq:profile_t4}
	\rho_z^{(4)} = \left[ 1 - \frac{1}{2} \left(\frac{z}{H^{(4)}}\right)^2 + \frac{1}{16} \left(\frac{z}{H^{(4)}}\right)^4 \right] 
\end{equation}
for the quartic form. The vertical heights $H^{(2)}$ and $H^{(4)}$ of the models are specified such that
the corresponding surface and midplane densities match the isothermal case, see Eq.~(\ref{eq:sigma}).
We obtain $H^{(2)} = 3/4 \sqrt{\pi} \, H$ and $H^{(4)} = 15/16 \sqrt{\pi/2} \, H$.
Fig.~\ref{fig:profiles} shows all three vertical density profiles in comparison.

\begin{figure}[htb]
	\centering
	\includegraphics[width=\columnwidth]{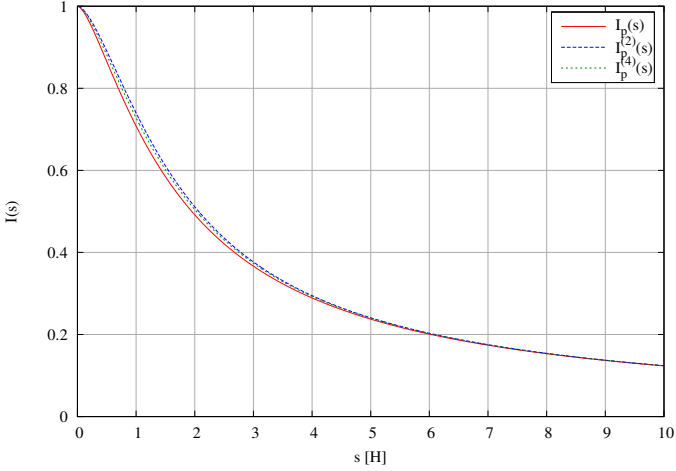}
	\caption{
		\label{fig:int_planet2}
		Force correction functions for different vertical density profiles. 
		$ I_\mathrm{p}(s) $ is the numerically calculated force correction function for the Gaussian vertical profile $ \rho_z $.
		$ I_\mathrm{p}^{(2)}(s) $ and $ I_\mathrm{p}^{(4)}(s) $ are the analytically calculated force correction
		functions for the parabolic vertical profile $ \rho_z^{(2)} $ and the fourth-order vertical profile $ \rho_z^{(4)} $.
	}
\end{figure}

These density stratifications are then used to calculate the vertically integrated force and to obtain the
corresponding force correction functions. Here, the vertical integrations extend to that $z_\mathrm{max}$ value where the density 
vanishes. For the two distributions we find $z_\mathrm{max}^{(2)} = \sqrt{2} \, H^{(2)}$ and $z_\mathrm{max}^{(4)} = 2 H^{(4)}$. 
For the second-order integral we find
\begin{equation}
	\label{eq:int_t2}
	I_\mathrm{p}^{(2)}(s)
	= \sqrt{1 + \frac{1}{2} c_2^2} \, - \frac{1}{2}c_2^2 \, \mbox{arcsinh} \left( \frac{\sqrt{2}}{c_2} \right) \,,
\end{equation}
and for the fourth-order integral
\begin{equation}
	\label{eq:int_t4}
	I_\mathrm{p}^{(4)}(s)
	= \frac{3c_4^2+8}{16} \sqrt{c_4^2+4} \, - \, \frac{c_4^2}{2} \left(1+ \frac{3 c_4^2}{16}\right)\,\mbox{arcsinh} \frac{2}{c_4} \,.
\end{equation}
Here, we defined $c_2 = s/H^{(2)}$ and  $c_4 = s/H^{(4)}$. As Fig.~\ref{fig:int_planet2} illustrates, for the 
unperturbed disk (without a planet)  these functions agree reasonably well with the Gaussian value
also for smaller separations $s$. For long distances all studied force correction functions 
($I_\mathrm{p}, I_\epsilon, I_\mathrm{p}^{(2)}$ and $I_\mathrm{p}^{(4)}$) approach each other.
These simpler profiles may be also useful in the study of self-gravitating disks.

\begin{acknowledgements}
	We thank Cl\'ement Baruteau and Sijme-Jan Paardekooper for useful discussions.
	Tobias M\"uller received financial support from the Carl-Zeiss-Stiftung. 
	Farzana Meru and Wilhelm Kley acknowledge the support of the German Research Foundation (DFG) through grant KL 650/8-2 
	within the Collaborative Research Group FOR 759: 
	{\it The formation of Planets: The Critical First Growth Phase}, and Farzana Meru is supported by the ETH Zurich Postdoctoral Fellowship Programm as well as by the Marie Curie Actions for People COFUND program.
	Most of the simulations were performed
	on the bwGRiD cluster in T\"ubingen, which is funded by the Ministry for Education and Research of Germany and
	the Ministry for Science, Research and Arts of the state Baden-W\"urttemberg, and the cluster of the
	Forschergruppe FOR 759 ``The Formation of Planets: The Critical First Growth Phase'' funded by
	the Deutsche Forschungsgemeinschaft. Finally, we thank the referee for the very constructive and
	and helpful comments.
\end{acknowledgements}

\bibliographystyle{aa}
\bibliography{mueller2012}

\end{document}